%% file: main_v2.tex
\documentclass[a4paper,12pt]{article}
\pdfoutput=1

\usepackage{jheppub}
\usepackage[sort&compress]{natbib}
\usepackage[utf8]{inputenc}
\usepackage[T1]{fontenc}

\usepackage{amsmath,amssymb,amsthm}
\usepackage{physics,mathtools}
\usepackage{float}
\usepackage{subfig}
\usepackage[svgnames]{xcolor}

\usepackage{comment}
\usepackage[normalem]{ulem}
\usepackage[font=small]{caption}

\hypersetup{linkcolor=DarkBlue,citecolor=DarkBlue, urlcolor=black} 
\makeatletter
\def\@fpheader{\vspace{1mm}}
\makeatother

\include{header.tex}

\def\zb{\overline{z}}
\def\Zb{\overline{Z}}

\newcommand{\xx}{|\vec{x}|}

\title{\LARGE{Bulkcone Singularities and Complex Geodesics}}  
\author{\hspace{0.02cm}Ignacio J. Araya,$^{\color{black}a}$}
\author{\hspace{0.02cm}Chantelle Esper,$^{\color{black}b}$} 
\author{\hspace{0.02cm}Yueke Jia,$^{\color{black}b}$} 
\author{\hspace{0.02cm}Manuela Kulaxizi,$^{\color{black}b,c}$} 
\author{Andrei Parnachev$^{\color{black}b}$\linebreak\vspace{-0.2cm}}

\affiliation[{\color{black}a}]{Departamento de F\'isica y Astronom\'ia, Facultad de Ciencias Exactas, \setlength{\parskip}{0pt}\newline\indent Universidad Andres Bello, Sazi\'e 2212, Piso 7, Santiago, Chile}

\affiliation[{\color{black}b}]{School of Mathematics and Hamilton Mathematics Institute, 
\setlength{\parskip}{0pt}\newline\indent Trinity College, Dublin 2, Ireland}

\affiliation[{\color{black}c}]{Division of Nuclear and Particle Physics, Department of Physics, 
\setlength{\parskip}{0pt}\newline\indent National and Kapodistrian University of Athens, Greece}

\abstract{Thermal correlators in holographic CFTs on a sphere exhibit bulk-cone singularities at points connected by null geodesics in the bulk. The operator product expansion analysis of the stress-tensor sector of the correlator shows that there are analogous singularities at spacelike separation for thermal CFTs on a plane. We show that these are associated with complex null geodesics. There is a phase transition between the real and complex spacelike geodesics underpinning this picture. We also provide a phase-shift calculation of the position of these generalised bulk-cone singularities.}

\begin{document}

\maketitle
\flushbottom
\newpage

\section{Introduction and Summary}
\label{sec.introduction}

The AdS/CFT correspondence \cite{Maldacena:1997re,Gubser:1998bc,Witten:1998qj} provides a useful laboratory for studying strongly interacting conformal field theories at finite temperature, which are dual to asymptotically AdS black holes \cite{Witten:1998zw}.
Correlators in finite-temperature holographic CFTs exhibit bulkcone singularities, which arise when the operator insertion points are connected by null geodesics in the bulk \cite{Hubeny:2006yu,Gary:2009ae,Maldacena:2015iua,Dodelson:2023nnr,Caron-Huot:2025hmkgm,Chen:2025jbf}.
Thermal correlators in holographic CFTs are obtained by solving the equation of motion for the dual bulk scalar; this is most easily done after Fourier transforming to momentum space.
Bulkcone singularities are dominated by the Regge behavior of the momentum-space correlator, which can be accessed using the WKB approximation, where the equation of motion reduces to that of a null geodesic.
Hence, the appearance of singularities at points connected by null geodesics is, in some sense, natural.

Bulkcone singularities have been studied for CFTs defined on  $\mathbb{R} \times S^{d-1}$ , which corresponds to the conformal boundary of asymptotically AdS spacetime in global coordinates.
Most physical questions, however, are formulated in the large-volume limit, where the radius of the sphere is taken to infinity, corresponding to the Poincaré patch of asymptotically AdS spacetime.
In this case, all null geodesics starting from the boundary fall into the horizon.
They may bounce off the black hole singularity and give rise to singularities in the two-sided correlator upon a suitable analytic continuation (see \cite{Kraus:2002iv,Fidkowski:2003nf,Festuccia:2005pi,Engelhardt:2014mea,Horowitz:2023ury,Parisini:2023nbd,DeClerck:2023fax,Ceplak:2024bja,Auzzi:2025sep,Afkhami-Jeddi:2025wra,Ceplak:2025dds,Dodelson:2025jff,Chakravarty:2025ncy,Jia:2025jbi,Kristjansen:2025xqo,AliAhmad:2026wem}
for related work).
However, at real time and real position, no bulkcone singularities have been observed.

One way to observe singularities in the two-sided correlator is to use the operator product expansion and focus on the contributions of multi-stress operators.
This part of the correlator, the stress-tensor sector, exhibits {\it bouncing singularities} (more precisely, the origin of the cut in the complex plane) at complex Euclidean times $\tau$ that correspond precisely to the points where the null geodesic, after bouncing off the black hole singularity, hits the boundaries of the eternal asymptotically AdS black hole.
The other part of the correlator comes from the contribution of double-trace operators.
One can argue that these contributions give a vanishing discontinuity across the cut, and hence it is the stress-tensor sector that is responsible for the singularity in the retarded Green’s function \cite{Afkhami-Jeddi:2025wra}.

To see the bouncing singularity in the simplest case, one analyzes the stress-tensor sector  at vanishing spatial separation and finds that the resulting series diverges for complex values of  $\tau=i t$.
One may then ask what happens if the same analysis is performed at t = 0 but at finite spatial separation, for a CFT in flat space   $\mathbb{R}^{d-1} \times \mathbb{R}$ .
As we show in this paper, the result is a singularity at a real value of the spatial separation.
Repeating this analysis for other values of $t$, one finds a curve of singularities.

As mentioned above, there is no null geodesic that connects such points on the flat boundary.
However, this statement comes with a caveat: there is no real null geodesic.
As it turns out, there are complex\footnote{Complex geodesics have been recently discussed in the de Sitter context in \cite{Chapman:2022mqd, Aalsma:2022eru}} (more precisely, purely imaginary) geodesics that correspond precisely to the curve $\vec x(t)$ where the stress-tensor sector diverges.
Hence, we find that complex geodesics give rise to bulkcone singularities in the stress-tensor sector. 

To analyze this in more detail, we consider the near-lightcone limit of finite-temperature CFT correlators, where only the leading-twist multi-stress tensors contribute.
In this limit the physics simplifies, since everything depends on a single variable $\alpha = x^- (x^+)^3 T^4$ (in $d=4$ spacetime dimensions).
In Appendix H of \cite{Ceplak:2025dds}, the stress-tensor sector was analyzed in this limit, and a real singularity was observed.
At the time this was puzzling, because real spacelike geodesics never approach a null one, and the corresponding correlator (at infinite conformal dimension) has an infinite radius of convergence \cite{Parnachev:2020fna}.
However, we now understand what must happen: one must look for a complex spacelike geodesic.
Indeed, such a complex geodesic exists and becomes thermodynamically favorable for sufficiently large values of $\alpha$.
It approaches a null geodesic at a critical value of $\alpha$, which governs the approach of the singularity curve $\vec{x}(t)$ to the lightcone.

\subsubsection*{Outline}

The paper is organised as follows. In section \ref{sec:boundry_ansatz} we use the OPE to extract the behaviour of the stress-tensor of the the thermal two-point function, both at vanishing and non-vanishing time separation.
We proceed to section \ref{sec:geodesics}, where we analyse imaginary null geodesics in the AdS black brane geometry, which are anchored at the boundary. In particular, we evaluate the coordinate and time displacement associated to such geodesics, compare with the position of the singularity observed in section \ref{sec:boundry_ansatz} and observe exact agreement. In section \ref{sec:thermal_two_point}, we investigate the presence of bulk-cone singularities associated to imaginary geodesics. In section \ref{sec:discussion} we summarise our findings and discuss open questions and directions for future work. Appendix \ref{app_bndry_anz} contains some technical details pertaining to the OPE computation of the correlator, including formulas which enable a direct matching of the CFT multi-stress-tensor OPE coefficients to those obtained from the holographic computation. Appendix \ref{app_small_t_coeffs} contains further details on the determination of the singularity from the OPE for spatial displacement varying with time. A discussion of spacelike geodesics and the light-like limit is presented in Appendix \ref{app_spacelike_geo}. Finally, Appendix \ref{app_exotic_bulk_geo} contemplates a complexified asymptotically AdS black-hole geometry where the imaginary null geodesics in question become real.

\section{Boundary Ansatz}
\label{sec:boundry_ansatz}

We are interested in computing holographic correlators at finite temperature, and in particular the two-point function of scalar operators, $\mathcal{O}$, with generic scaling dimension $\Delta$ at finite temperature $\beta = 1/T$. In holographic theories the correlator can be decomposed as follows,
\begin{equation}
\label{eq:defG}
    \langle \mathcal{O}(t, \vec{x}) \mathcal{O}(0,0)\rangle_\beta = G(t, \vec{x}) = G_T(t, \vec{x}) + G_{[\mathcal{O} \mathcal{O}]}(t, \vec{x})\,.
\end{equation}
Here $G_T(t,\vec{x})$ denotes the stress-tensor sector of the correlator which is determined by the multi-stress tensor contributions in the OPE. Similarly, $G_{[\mathcal{O} \mathcal{O}]}(t, \vec{x})$ denotes the sector composed of double-trace contributions. In what follows we will restrict the analysis to four-dimensional $CFT$s dual to $AdS_5$ gravity. The methodology can be generalised to $CFT_d$ in general even dimensions, but we focus on $d=4$ for simplicity. 

The holographic computation of the two-point function at finite temperature requires studying scalar field fluctuations around the AdS-Schwarzschild black brane background given by,
\begin{equation}
    ds^2 = -\frac{r^2}{R_{AdS}^2} f(r) dt^2 + \frac{ R_{AdS}^2}{r^2 f(r)} dr^2+ \frac{r^2}{R_{AdS}^2} d\vec{x}^2\,, \quad f(r) = 1-\frac{\mu}{r^4}\,.
    \label{eq:bcknd}
\end{equation}
The horizon is located at $r = \mu^{1/4}$, and the temperature of the geometry is related to the parameter $\mu$ as follows,
\begin{equation}
    \mu = (\pi T R_{AdS}^2)^4\,.
    \label{eq:defmuT}
\end{equation}
We set $R_{AdS} =1$ and $\mu =1$ such that $\beta = \frac{1}{T}=\pi$ throughout the rest of the paper for convenience.  

In this section, we employ the near boundary ansatz introduced in \cite{Fitzpatrick:2019zqz}, to compute $G_T(\tau, \vec{x})$ in Euclidean signature. We can then Wick rotate to Lorentzian signature without crossing any branch cuts since we will be considering spacelike separation $|\vec{x}|^2 > t^2$. The method is reviewed below. The starting point is the equation of motion for a minimally coupled scalar field $\phi(\tau, \vec{x})$, dual to $\mathcal{O}$, in the background given by (\ref{eq:bcknd}). 

\begin{equation}
   \left( \Box- \Delta(\Delta-4) \right)\phi(r,\tau, \vec{x}) = 0\,,
   \label{eq:KGEq}
\end{equation}

We introduce new coordinates $\rho, \lambda$, that isolate the leading asymptotic behaviour near the
boundary,\footnote{These coordinates are different than the coordinates used in \cite{Fitzpatrick:2019zqz}}
\begin{equation}
    \rho = r \tau\,, \quad \lambda^2 = 1+ r^2 \left(|\vec{x}|^2+\tau^2 \right)\,,
   \label{eq:defrholambda} 
\end{equation}
and factor out the pure AdS solution,
\begin{equation}
    \phi(r,\rho, \lambda) = \left( \frac{r}{\lambda^2}\right)^\Delta\psi(r, \rho, \lambda)\,.
    \label{eq:phitopsi}
\end{equation}
The equation (\ref{eq:KGEq}) reduces to,
\begin{equation}
    \mathcal{D} \psi(r,\rho, \lambda) = 0\,,
    \label{eq:diff_op_psi}
\end{equation}
where the differential operator $\mathcal{D}$ can be found in appendix \ref{app_bndry_anz}.

The correlator $G_T$ is then determined as follows \cite{Fitzpatrick:2019zqz},
\begin{equation}
    G_T(\tau, |\vec{x}|) = \frac{1}{\left( |\vec{x}|^2+\tau^2\right)^{\Delta}}\lim_{r \to \infty} \psi(r, \tau, |\vec{x}|)\,,
    \label{eq:GTrelpsi}
\end{equation}
where $\psi$ is given by the following series expansion, \footnote{We have chosen to include a factor of $(-1)^j$ to absorb the effects from Wick rotation}
\begin{equation}
     \psi(r, \rho, \lambda)  = \sum_{i=0}^\infty \sum_{j=0}^i \sum_{k=-i}^{2i-j} \frac{\rho^{2j} \lambda^{2k}}{r^{4i}} (-1)^j a^i_{j, k}\,, \quad a^0_{0, 0}=1\,.
     \label{eq:psi_anz}
\end{equation}
The stress-tensor sector of the thermal correlator can be expressed in terms of the coefficients $a^i_{j, k}$ as follows,
\begin{equation}
    G_T(t, |\vec{x}|) = \frac{1}{(|\vec{x}|^2-t^2)^{\Delta}} \sum_{i=0}^\infty \sum_{j=0}^i t^{2j} \left( |\vec{x}|^2-t^2\right)^{2i-j} a^i_{j, 2i-j}\,,
\label{eq:corr_nonstand_gen}
\end{equation}
where we have Wick rotated back to Lorentzian signature, since the singularities that we will find at finite $|\vec{x}|$ are located at real Lorentzian time. We will   henceforth be considering the correlator in $G_T(t, |\vec{x}|)$ Lorenztian signature.

\subsection{Singularity at Zero Time.}
\label{sec:bndry_anz_t0}

\begin{figure}
    \centering
    \includegraphics[width=0.9
    \linewidth]{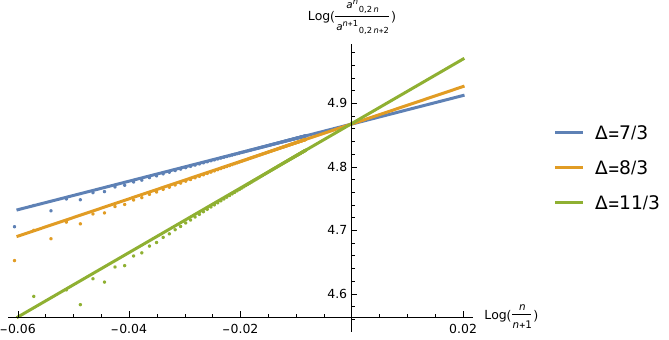}
    \caption{Log-log plots of the ratio of successive coefficients $a^n_{0, 2n}$ for various values of $\Delta$, alongside the linear fits of the data for large $n$ following (\ref{eq:t_0_asymp_coeff}). Notice that the $y$-intercept, given by $\log{B}$, appears independent of $\Delta$. We see that the fit to (\ref{eq:t_0_asymp_coeff}) is precise, with the fit parameters for various $\Delta$ displaying standard deviations of $\mathcal{O}(10^{-5})$. For smaller $n$, corresponding to more negative values of $\log{(n/n+1)}$, the data ceases to fit (\ref{eq:t_0_asymp_coeff}) as expected.}
    \label{fig:t_0_coeff_log}
\end{figure}

We can predict the existence and position of non-analytic features in $G_T(t, |\vec{x}|)$ by analysing when the series expansion (\ref{eq:corr_nonstand_gen}) fails to converge. We begin by investigating the singularities of the correlator at $t=0$, for which (\ref{eq:corr_nonstand_gen}) simplifies to,
\begin{equation}
    G_T(t=0, |\vec{x}|) = \frac{1}{|\vec{x}|^{2 \Delta}} \sum_{n=0}^\infty |\vec{x}|^{4n} a^n_{0, 2n}\,.
    \label{eq:corr_t_0_series}
\end{equation}
The asymptotic behaviour of the coefficients $a^n_{0, 2n}$ determine the convergence of the series representation (\ref{eq:corr_t_0_series}) of the correlator.
In the limit of large n, the coefficients $a^n_{0, 2n}$ can be approximated by,
\begin{equation}
  \lim_{n \to \infty}  a^n_{0, 2n} = \frac{A(\Delta) n^{2 \Delta - 5/2}}{B^{4n}}\,,
  \label{eq:t_0_asymp_coeff}
\end{equation}
where $B$ is determined numerically for several values of $\Delta$ to be $B \approx 3.37(7)$. 
Here $A(\Delta)$ is a complicated function of $\Delta$, the specifics of which are not important for the results in this paper.
One can compute $B$ by performing linear regression on $a^n_{0, 2n}$ as follows \footnote{This allows for a clear separation between the effect of $B$ and the rest of the asymptotic form (\ref{eq:t_0_asymp_coeff}). }, 
\begin{equation}
    \log{\left( \frac{a^{n}_{0, 2n}}{a^{n+1}_{0, 2n+2}}\right) = \left(2\Delta - \frac{5}{2}\right)\log{\left(\frac{n}{n+1} \right)}} + 4 \log{B}\,.
    \label{eq:logcoeff}
\end{equation}
In figure \ref{fig:t_0_coeff_log}, we show how the coefficients $a^n_{0, 2n}$ converge to the asymptotic behaviour (\ref{eq:t_0_asymp_coeff}) for various values of $\Delta$. The subleading corrections to (\ref{eq:t_0_asymp_coeff}) become more important for larger $\Delta$, as discussed in \cite{Ceplak:2024bja}.

The asymptotic form of $a^n_{0, 2n}$ can be inserted into (\ref{eq:corr_t_0_series}), and the sum approximated by an integral to find,
\begin{equation}
\begin{aligned}
      G_T(t=0, |\vec{x}|) & \approx \frac{A(\Delta)}{|\vec{x}|^{2\Delta}} \int_0^\infty dn \left(\frac{|\vec{x}|^4}{B^4} \right)^n n^{2\Delta - \frac{5}{2}} \\
      & = \frac{A(\Delta) \Gamma\left(2\Delta-\frac{3}{2}\right)}{|\vec{x}|^{2\Delta}} \left( - \log{\left( \frac{|\vec{x}|^4}{B^4}\right)}\right)^{\frac{3}{2} - 2\Delta}\\
     & \xrightarrow[|\vec{x}| \approx B^4]{} \frac{A(\Delta)\Gamma\left(2\Delta - \frac{3}{2} \right)}{4^{2\Delta - 3/2} B^{3/2}}\left(\frac{1}{B-|\vec{x}|}\right)^{2\Delta -\frac{3}{2}}\,.
     \label{eq:corr_t0_integral}
\end{aligned}
\end{equation}
Hence, we see that the stress tensor sector of the correlator at $t=0$ has a divergence at,
\begin{equation}
    |\vec{x}| = B \approx 3.37(7)\,.
    \label{eq:t0_sing_predict}
\end{equation}

Notice that the singularity predicted for $\tau = t = 0$, (\ref{eq:t0_sing_predict}), lies outside of the minimum radius of convergence of the OPE, which is determined by the KMS pole to be,
\begin{equation}
    \sqrt{|\vec{x}|^2 + \tau^2} = \pi\,.
\end{equation}

\subsection{Singularities for Finite Time}
\label{sec:bndry_anz_xt}
Having seen the singularity in the correlator for $t=0$, we want to investigate the possible singularities at generic spacelike values of $(t, |\vec{x}|)$. To do so, we reorganise the series expansion in (\ref{eq:corr_nonstand_gen}) as a power series around the lightcone. 
\begin{equation}
    \begin{aligned}
         G_T(t,|\vec{x}|) =& \frac{1}{(|\vec{x}|^2-t^2)^{\Delta}} \sum_{i=0}^\infty \sum_{j=0}^i t^{2j} \left( |\vec{x}|^2-t^2\right)^{2i-j} a^i_{j, 2i-j}\\
         =&  \frac{1}{(|\vec{x}|^2-t^2)^{\Delta}} \sum_{i=0}^\infty \sum_{j=0}^i y^{4i-2j}  t^{4i}  \ a^i_{j, 2i-j}\,, \quad y^2 = \frac{|\vec{x}|^2}{t^2}-1\\
         =&  \frac{1}{(|\vec{x}|^2-t^2)^{\Delta}} \sum_{i=0}^\infty \sum_{n=i}^{2i} y^{2n}t^{4i} \ a^i_{2i-n, n}\\
         =&  \frac{1}{(|\vec{x}|^2-t^2)^{\Delta}} \sum_{n=0}^\infty y^{2n} \Lambda_n(t) \,,
    \end{aligned}
    \label{eq:corr_lc_seris}
\end{equation}
where we have exchanged the bounds of summation for indices $i, n$ and have defined the series coefficients $\Lambda_n(t)$ as follows,
 \begin{equation}
        \Lambda_n(t) = \sum_{i=\lceil n/2 \rceil}^{n} t^{4i} \  a^i_{2i-n, n} \,,
         \label{eq:corr_sing_t0}
 \end{equation}
where $\lceil x \rceil$ denotes the smallest integer $m$ such that $m\geq x$.
Much like in section \ref{sec:bndry_anz_t0}, the asymptotic behaviour of the series coefficients $\Lambda_n(t)$ can identify possible singularities along the path of constant $t$.

\begin{figure}
    \centering
    \includegraphics[width=0.9\linewidth]{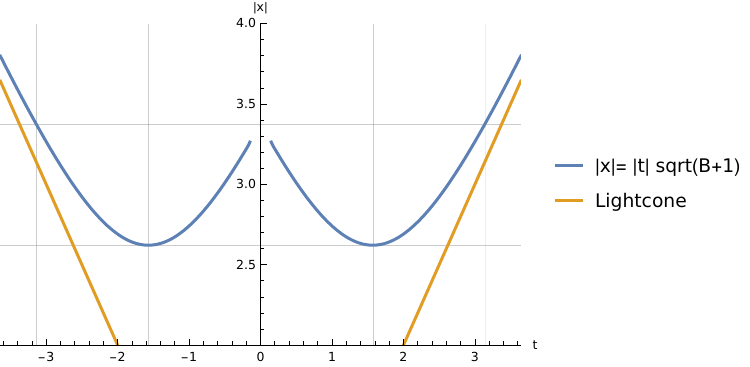}
    \caption{Plot of the position of singularity $|\vec{x}|$ as a function of $t$, as predicted by (\ref{eq:corr_sing_t_val}) for $\Delta = 7/3$, alongside the lightcone $|\vec{x}| = t$. Gridlines are shown for $t = \pm \frac{\pi}{2}, \pm \pi$ and $|\vec{x}| = \frac{\sqrt{\pi}}{2}\frac{\Gamma(1/4)}{\Gamma(3/4)}, 3.37$. The upper $|\vec{x}|$ gridline is the singularity position in $|\vec{x}|$ when $t=0$. The gridlines show that the singularity curve is symmetric about $t = \frac{\pi}{2}$. The singularity curve asymptotes from above to the lightcone at large $t$. Values of $|t| < 0.2$ are excluded due to a loss in accuracy for small $t$, see appendix \ref{app_small_t_coeffs} for further details.}
    \label{fig:x_t_sing}
\end{figure}

The coefficients $\Lambda_n(t)$ approach the same asymptotic form as in (\ref{eq:t_0_asymp_coeff}) for large $n$,
 \begin{equation}
   \lim_{n \to \infty}  \Lambda_n (t) = \frac{A(\Delta, t) n^{2 \Delta -5/2}}{B(t)^n}\,,
   \label{eq:t0_sing_ansatz}
    \end{equation}
    where $A(\Delta, t)$ is an undetermined function.
We approximate $G_T(t, |\vec{x}|)$ by substituting (\ref{eq:t0_sing_ansatz}) into (\ref{eq:corr_lc_seris}) and exchanging the sum with an integral as in (\ref{eq:corr_t0_integral}),
\begin{equation}
\begin{aligned}
      G_T(t, |\vec{x}|) & \approx 
       \frac{A(\Delta, t)\Gamma\left( 2 \Delta - \frac{3}{2}\right)}{\left(|\vec{x}|^2-t^2 \right)^{\Delta}} \left(- \log{\left( \frac{|\vec{x}|^2}{t^2 B(t)}-\frac{1}{B(t)} \right)}\right)^{\frac{3}{2}-2\Delta}\\
       & \approx \frac{A(\Delta, t) \Gamma\left( 2 \Delta - \frac{3}{2}\right)}{t^{3/2} 2^{2\Delta - 3/2}} \frac{B(t)^{\Delta - 3/2}}{(B(t)+1)^{\Delta - 3/4}} \left( \frac{1}{|\vec{x}| - t\sqrt{1+B(t)}}\right)^{2\Delta - \frac{3}{2}}\,.
      \label{eq:corr_sing_form}
\end{aligned}
\end{equation}
Hence, the correlator exhibits a singularity at, \footnote{Since the series coefficients $\Lambda_n(t)$ are invariant under Wick rotation, we can immediately conclude from (\ref{eq:corr_sing_t_val}) that real values of Euclidean time $\tau$ correspond to singularities in imaginary $|\vec{x}|$, which we do not allow.}
\begin{equation}
        |\vec{x}| = \pm t \sqrt{B(t)+1}\,.
        \label{eq:corr_sing_t_val}
\end{equation}
The position of the singularity curve $|\vec{x}| = |\vec{x}|(t)$ is depicted in figure \ref{fig:x_t_sing}.  To generate the plot, we extract $B(t)$ from $\Lambda_n(t)$ with largest $n$ for a given value of t. This produces the smooth curve of the figure.  Alternatively, one could fit the coefficients $\Lambda_n(t)$ for large $n$ as in figure \ref{fig:t_0_coeff_log} for specific values of t. This would result in a more accurate estimate of $B(t)$, however the difference is very small.\footnote{For instance, the relative error between fitting asymptotic behaviour and considering only the most asymptotic coefficient values, for $t=\pi/2$, is only $0.07$\%.} The convergence of $\Lambda_n(t)$ to the asymptotic form (\ref{eq:t0_sing_ansatz}) is similar to that of the $t=0$ coefficients $a^n_{0, 2n}$ except for $t < 0.2$, where convergence of the numerics drastically decreases. Formally, this is due to the definition of the expansion parameter $y$ around the lightcone, which diverges at $t=0$. We can remedy this by choosing a different expansion parameter in the series. The interested reader can view the new series expansion and the resulting singularity curve in detail in appendix \ref{app_small_t_coeffs}, which recovers the small t behaviour.

\begin{figure}
    \centering
    \includegraphics[width=0.7\linewidth]{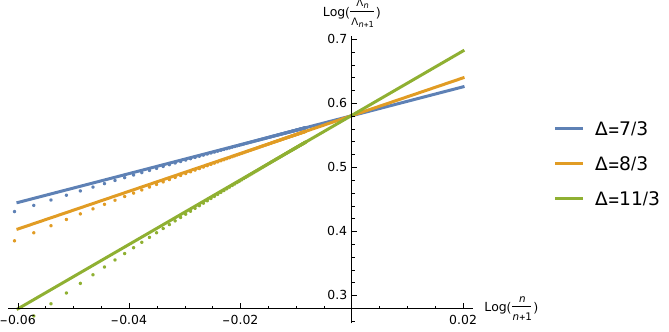}
    \caption{Log-Log plot of the ratio of successive coefficients $\Lambda_n$ for $t=\pi/2$, for various values of $\Delta$. Shown also are the linear fits for large $n$.}
    \label{fig:t_pi2_coeffs_log}
\end{figure}

It is useful to explicitly compute the singularity position (\ref{eq:corr_sing_t_val}) for $t = \pi/2$, corresponding to the minimum of the curve in figure \ref{fig:x_t_sing}. The corresponding value for $|\vec{x}|$ can be compared with an analytic value which we compute in section \ref{sec:geodesics}.
Figure \ref{fig:t_pi2_coeffs_log} displays the convergence of the series coefficients $\Lambda_n(t=\pi/2)$ to the asymptotic form (\ref{eq:t0_sing_ansatz}). Similar to the analysis for $t=0$, convergence is slower as $\Delta$ is increased. 
Fitting $\Lambda_n$ for large $n$ with $\Delta = 7/3$ results in the following estimate for $|\vec{x}|(t=\pi/2)$, 
\begin{equation}
    |\vec{x}|\left(t=\frac{\pi}{2}\right) \approx 2.622(4)\,.
    \label{eq:t_pi2_sing_predic}
\end{equation}

\section{Geodesics in the Planar Black Hole Background}
\label{sec:geodesics}
In this section we investigate whether the positions of the singularities predicted numerically in section \ref{sec:boundry_ansatz}, correspond to null geodesics in the asymptotically $AdS_5$ black brane background (\ref{eq:bcknd}). However, no real null geodesics, which return to the boundary, exist in this background. Hence, we explore whether complex null geodesics could explain such singularities. \footnote{We define a complex geodesic as one in which the turning point is not the real turning point closest to the boundary. The geodesics that we will consider are described by a bulk contour that lies outside of the physical regime of the bulk radial coordinate. }

The asymptotically $AdS_5$ black brane metric is given by,
\begin{equation}\label{BBAdS5}
    ds^2 = - \frac{(1-u^2)}{u}dt^2 + \frac{1}{u}d\vec{x}^2 + \frac{1}{4u^2 (1-u^2)}du^2\,,
\end{equation}
where we use $u = r^{-2}$ as the bulk radial coordinate. 
The conformal boundary corresponds to $u=0$, and the horizon is given by $u=1$. 
To match the boundary ansatz computation, we will exploit the spatial spherical symmetry to define $|\vec{x}| = \sqrt{x^2 + y^2 + z^2}$ as the spatial displacement. 
The definitions of the conserved quantities are given by,
\begin{equation}
    |\dot{\vec{x}}| = |\vec{P}| u\,, \quad \dot{t} = \frac{E u}{(1-u^2)} = \frac{b |\vec{P}| u}{(1-u^2)}\,.
    \label{eq:geo_derv_def}
\end{equation}
where the impact parameter $b$ is given by,
\begin{equation}
    b=E/|\vec{P}|\,.
    \label{eq:defbnull}
\end{equation}

\begin{figure}
    \centering
\includegraphics[width=0.5\linewidth]{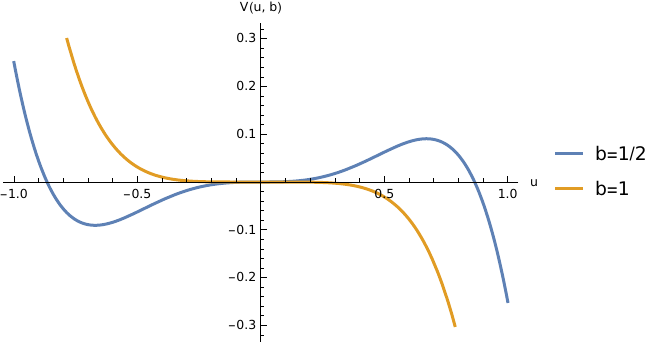}
    \caption{Effective potential for null geodesics, shown for $b=1/2$ (blue) and $b=1$ (orange). We have set $|P| = 1/\sqrt{2}$.}
    \label{fig:geo_potential}
\end{figure}

The equation of motion for null geodesics is given by,
\begin{equation}
    \frac{1}{2}\dot{u}^2 = - V(u, b)\,, \quad V(u, b) = 2 |P|^2 u^3(1-b^2-u^2)\,.
    \label{eq:nullgeqandpotential}
\end{equation}
Note the effective energy for this 1D classical system is zero. The effective potential, $V(u, b)$, for the geodesic with $b^2<1$ is shown in blue in figure \ref{fig:geo_potential}. It is classically forbidden for a particle in such a potential starting at the boundary $u=0$ to move within the physical bulk $u>0$. 

Allowing $u<0$, we can create solutions where a particle will return to the boundary (see figure \ref{fig:geo_potential}). Given that the standard radial coordinate $r = u^{-1/2}$, this corresponds to particles moving in imaginary $r$. These are the complex geodesics that we referred to at the beginning of this section.
The turning point for the complex geodesic path in the bulk is determined by the negative solution to the following equation $V(u, b) = 0$, which is,
\begin{equation}
    u = - \sqrt{1-b^2} \equiv - u_*\,.
    \label{eq:turning_pt}
\end{equation}

We restrict ourselves to the regime $b^2<1$, for which the geodesic turning points are real in the coordinate $u$. Outside this range, where the turning points in $u$ are complex, the spatial displacement $|\vec{X}|$ for any complex geodesic that returns to the same boundary is also complex, which we neglect.

\begin{figure}
    \centering
    \includegraphics[width=0.5\linewidth]{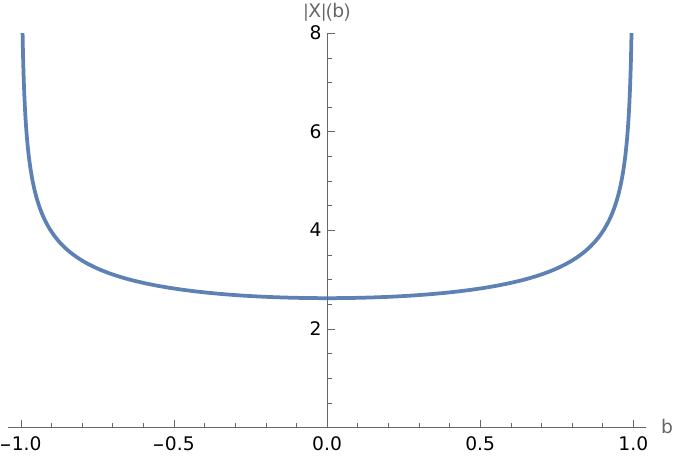}
    \caption{Spatial displacement for null geodesic $|\vec{X}|(b)$ as a function of the impact parameter $b$. }
    \label{fig:geo_x}
\end{figure}

The time and spatial displacement for the complex null geodesic are defined as follows,
\begin{equation}
    \begin{aligned}
        T(b) & = b \int_0^{-u_{*}} du\frac{u}{(1-u^2)\sqrt{u^3(u^2-u_{*}^2)}}\,, \\
    |\vec{X}|(b) & = \int_0^{-u_{*}} du \frac{u}{\sqrt{- u^3 \left(u_{*}^2-u^2\right)}}\,.
    \end{aligned}
    \label{eq:null_geo_t_x_def}
\end{equation}
The spatial displacement $|\vec{X}|$ evaluates to,
\begin{equation}
    |\vec{X}|(b)= \frac{2 \sqrt{\pi}}{(1-b^2)^{1/4}} \frac{\Gamma{(5/4)}}{\Gamma{(3/4)}} \,.
        \label{eq:null_geo_x}
\end{equation}
A plot of (\ref{eq:null_geo_x}) is shown in figure \ref{fig:geo_x}. The minimum of $|\vec{X}|(b)$ is attained at $b=0$ and is given by,
\begin{equation}
    |\vec{X}|(b=0) = \frac{2\sqrt{\pi} \Gamma(5/4)}{\Gamma(3/4)}\,.
    \label{eq:Xforbzero}
\end{equation}

\begin{figure}
    \centering
    \includegraphics[width=0.5\linewidth]{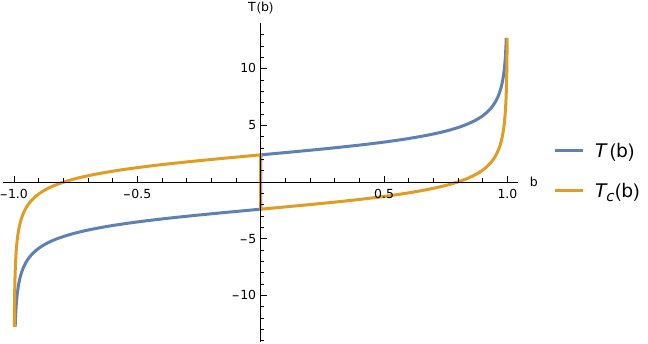}
    \caption{Time displacement $T(b)$ (\ref{eq:geo_t_1}) (blue) and its analytic continuation $T_c(b)$ (\ref{eq:geo_t_2}) (orange) for complex null geodesics as a function of the impact parameter b. While neither of the curves $T(b)$, $T_c(b)$ are continuous at $b=0$, it is possible to piecewise define continuous functions of $b$, given by $T_\pm(b)$ in (\ref{eq:null_geo_T_b_plus}), (\ref{eq:null_geo_T_b_minus}).}
    \label{fig:geo_t_b}
\end{figure}

The time displacement $T(b)$ can be evaluated as follows,
\begin{equation}
    T(b) = \frac{\sqrt{\pi}}{2}\frac{\Gamma(1/4)}{\Gamma(3/4)}\frac{b}{(1-b^2)^{1/4}}{}_2F_1\left(1, \frac{1}{4}, \frac{3}{4}, 1-b^2\right)\,.
    \label{eq:geo_t_1}
\end{equation}
This expression naively produces two discontinuous curves, shown in blue in figure \ref{fig:geo_t_b}. It is clear from the figure that (\ref{eq:geo_t_1})
does not have solutions for $T(b) \in (-\pi/2, \pi/2)$. Notice, however, that the hypergeometric function in (\ref{eq:geo_t_1}) exhibits a branch cut, starting at $b=0$ and extending for all $b^2<0$. To analytically continue around $b=0$, we use the following formula,
\begin{equation}
\begin{aligned}
     {}_2F_1\left(a, b, \right.
     &\left. c,\,
      1-b^2 e^{2\pi i} \right)  =  e^{2\pi i(a + b -c )}{}_2F_1\left(a, b, c, 1-b^2\right) + \\
     & + \frac{2\pi i e^{\pi i(a+b-c)}\Gamma(c)}{\Gamma(c-a)\Gamma(c-b)\Gamma(a+b-c+1)}{}_2F_1\left(a, b, a+b-c+1, b^2 \right)\,.
     \label{eq:2f1_cont}
\end{aligned}
\end{equation}
The analytically continued time displacement is given by,
\begin{equation}
\small
    T_c(b) = - \frac{\sqrt{\pi}}{2}\frac{\Gamma(1/4)}{\Gamma(3/4)}\frac{b}{(1-b^2)^{1/4}} \left({}_2F_1\left( 1, \frac{1}{4}, \frac{3}{4}, 1-b^2\right) - {}_2F_1\left(1, \frac{1}{4}, \frac{3}{2}, b^2 \right)  \right)\,.
    \label{eq:geo_t_2}
\end{equation}
We can combine (\ref{eq:geo_t_1}) for $b>0$ and the analytically continued expression for $b<0$ to obtain a continuous null geodesic time displacement $T_+(b)$, given by,
\begin{equation}
     T_+(b) = \frac{\left( 2\pi\right)^{3/2}}{\Gamma\left(-1/4 \right)^2} \left(4 \frac{b}{(1-b^2)^{1/4}} -{\sgn{b}}\,\beta\left[|b|^2,\frac{1}{2},\frac{3}{4}\right]\right) + \frac{\pi}{2}\,.
     \label{eq:null_geo_T_b_plus}
\end{equation} 
Here we express the result in terms of an incomplete beta function, by using standard identities for hypergeometric function.
This way we obtain the upper continuous curve in figure \ref{fig:geo_t_b}. To produce the lower continuous curve in figure \ref{fig:geo_t_b}, we start with (\ref{eq:geo_t_1}) for $b<0$, and analytically continue for $b>0$. The result, which we call $T_-(b)$, is given by,
\begin{equation}
     T_-(b) = \frac{\left( 2\pi\right)^{3/2}}{\Gamma\left(-1/4 \right)^2} \left(4 \frac{b}{(1-b^2)^{1/4}} -\sgn{b}\beta\left[|b|^2,\frac{1}{2},\frac{3}{4}\right]\right) - \frac{\pi}{2}\,.
     \label{eq:null_geo_T_b_minus}
\end{equation} 

\begin{figure}
    \centering
    \includegraphics[width=0.9\linewidth]{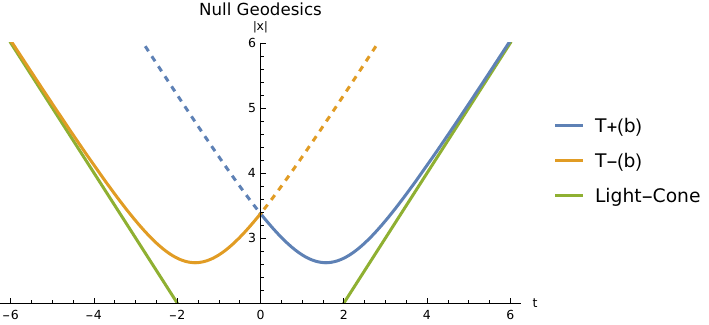}
    \caption{Parametric plot of the continuous null geodesics $(T_+(b), X(b))$ and $(T_-(b), X(b))$. The analytically continued portions of both $T_+(b)$ and $T_-(b)$ account for the singularities predicted by the boundary ansatz for $t \in [-\pi/2, \pi/2]$ in figure \ref{fig:x_t_sing}. The dashed lines cannot be determined by the methods of section \ref{sec:boundry_ansatz}.}
    \label{fig:null_geo_full}
\end{figure}

We now compare the singularity curves as predicted in section \ref{sec:boundry_ansatz} to the complex null geodesics shown in figure \ref{fig:null_geo_full}. A parametric plot of the null geodesics described by $T_+(b)$ and $T_-(b)$ is given in figure \ref{fig:null_geo_full}.  The function composed of the solid blue and orange curves precisely match those of figure \ref{fig:x_t_sing}\footnote{The relative error is of order $0.01$\%, and decreases for large $T_{\pm}$}.
The dashed lines in figure \ref{fig:null_geo_full} cannot be replicated via the methods of section \ref{sec:boundry_ansatz}, since one can only determine the position of the singularity that is closest to the lightcone.

In section \ref{sec:bndry_anz_t0}, we computed the position of the singularity at $t=0$ given by (\ref{eq:t0_sing_predict}). To find the value of $|\vec{X}|$ at $T_{\pm} = 0$ for the null geodesics, we must solve the following equation,
\begin{equation}
    \beta\left(1-\left( \frac{\Gamma(1/4)}{\Gamma(3/4)}\frac{\sqrt{\pi}}{2|\vec{X}|}\right)^4 , \frac{1}{2}, \frac{3}{4}
    \right) = \frac{8 |\vec{X}|}{\sqrt{\pi}}\frac{\Gamma(3/4)}{\Gamma(1/4)}\sqrt{1-\left( \frac{\Gamma(1/4)}{\Gamma(3/4)}\frac{\sqrt{\pi}}{2|\vec{X}|}\right)^4} - \frac{\Gamma(-1/4)^2}{4\sqrt{2\pi}}\,,
    \label{eq:X_Tzero_eq} 
\end{equation}
which can be solved numerically to arbitrary precision, resulting in,
\begin{equation}
    T_{\pm} = 0 \implies |\vec{X}| \approx 3.3763\,.
    \label{eq:X_Tzero_value}
\end{equation}
The relative difference between the null geodesic value above and (\ref{eq:t0_sing_predict}) is $0.01$\%.

Recall that the position of the singularity in $|\vec{x}|$ for $t = \pi/2$ was computed in (\ref{eq:t_pi2_sing_predic}). For the null geodesics, the corresponding value of $|\vec{X}|$ for $T_{\pm} = \pm \pi/2$ is analytic and given by,
\begin{equation}
    T_{\pm} = \pm \frac{\pi}{2} \implies |\vec{X}| = \frac{\sqrt{\pi}}{2}\frac{\Gamma(1/4)}{\Gamma(3/4)}\,.
    \label{eq:X_Tpiovertwo_value}
\end{equation}
The relative difference between this and (\ref{eq:t_pi2_sing_predic}) is $0.02\%$. 

\section{Thermal Two-Point Function and Singularity}\label{sec:thermal_two_point}

In this section, we would like to examine the existence of singularities analytically.
For reasons of convenience, the starting point will be the asymptotically AdS black
hole instead of the black brane, to which we will turn to in the end.

The $(d+1)$-dimensional AdS-Schwarzschild metric is given by,
\begin{equation}\label{eq:adssch}
ds^2 = -f\,dt^2 + f^{-1}dr^2 + r^2 d\Omega^2
\end{equation}
where,
\begin{equation}\label{eq:sphere metric}
d\Omega^2 = d\theta^2 + \sin^2\theta\, d\Omega^2_{d-2}\,,
\end{equation}
and,
\begin{equation}\label{eq:metricfunctionbh}
f = 1 + \frac{r^2}{R^2} - \frac{\mu}{r^{d-2}}\,, \qquad
\mu \equiv \left(\frac{d-1}{16\pi}\,\Omega_{d-1}\right)^{-1} G_N M\,.
\end{equation}
$R$ denotes both the AdS radius and the radius of the $(d-2)$-dimensional sphere.
The Hawking temperature $T_H$ is \cite{Witten:1998zw}, 
\begin{equation}\label{eq:HawkingT}
T_H = \frac{1}{\beta} = \frac{d\, r_H^2 + (d-2)R^2}{4\pi R^2 r_H}\,,
\end{equation}
where $r_H$ denotes the position of the horizon given by,
\begin{equation}\label{eq:rHdef}
f(r=r_H) = 0\,.
\end{equation}
To conform to the notations of the previous sections, we will also use the radial
coordinate $u$ defined as,
\begin{equation}\label{eq:rtou}
u = \frac{1}{r^2}\,,
\end{equation}
which allows us to express the AdS black hole metric as follows (we set $\mu=R=1$
for convenience),
\begin{equation}\label{eq:BHmetricu}
ds^2 = -f(u)\,dt^2 + \frac{1}{4u^3 f(u)}\,du^2 + \frac{1}{u}\,d\Omega^2_{d-1}\,,
\qquad f(u) = \frac{1}{u} + 1 - u^{\frac{d-2}{2}}\,.
\end{equation}

Assuming that the Wightman function in momentum space, $G_+(\omega,\ell)$, is known,
its coordinate space representation can be computed by a Fourier transform:
\begin{equation}\label{eq:Gplus1}
G_+(t,\theta) = \sum_{l=0}^{\infty}
\frac{\left(l+\frac{d-2}{2}\right)\Gamma\!\left(\frac{d}{2}\right)}{(d-2)\pi^{d/2}}\,
C_l^{\frac{d-2}{2}}(\cos\theta)
\int_{-\infty}^{\infty}\frac{d\omega}{2\pi}\, e^{-i\omega t}\, G_+(\omega,l)\,,
\end{equation}
where $C_l^{\frac{d-2}{2}}(\cos\theta)$ are Gegenbauer polynomials and the angular
variable $\theta$ lies in the range $0<\theta<\pi$.

Here we are interested in the eikonal regime, where $(\omega,\ell)$ are large in units of
temperature but the ratio remains finite. For convenience, we set
$l+\frac{d}{2}-1 = k$ with $k\gg1$ and use the following approximation for the
Gegenbauer polynomials,
\begin{equation}\label{eq:Gegenbauer}
C^{\frac{d-2}{2}}_{\ell=k-\frac{d-2}{2}}(\cos\theta) \approx
\frac{1}{2}\left(\frac{2}{k}\right)^{(4-d)/2}
\frac{e^{-i\pi(d-2)/4}\, e^{i\pi\frac{d-2}{2}\lfloor\frac{\theta}{\pi}\rfloor}}
{\Gamma\!\left(\frac{d-2}{2}\right)\left(|\sin\theta|\right)^{(d-2)/2}}
\left(e^{ik\theta} + e^{i\pi(d-2)/2}\, e^{-ik\theta}\right).
\end{equation}
Here $\lfloor x\rfloor$ denotes the largest integer $n$ such that $n\leqslant x$\footnote{We have defined
$e^{i\pi\frac{d-2}{2}\lfloor\frac{\theta}{\pi}\rfloor}\equiv1$ for $\theta\in(0,\pi)$
in \eqref{eq:Gegenbauer}.}.
Substituting \eqref{eq:Gegenbauer} into \eqref{eq:Gplus1} and replacing the summation
over $l$ by integration over $k$, allows us to write the position space two-point
function as follows,
\begin{equation}\label{eq:Gplus3}
G_+(t,\theta) = g_+(t,\theta) + e^{i\pi(d-2)/2}\, g_+(t,-\theta)\,,
\end{equation}
with $g_+(t,\theta)$ defined as,
\begin{equation}\label{eq:gplusdef}
g_+(t,\theta) = \frac{e^{-i\pi(d-2)/4}\, e^{i\pi\frac{d-2}{2}\lfloor\frac{\theta}{\pi}\rfloor}}
{(2\pi)^{d/2+1}\left(|\sin\theta|\right)^{d/2-1}}
\int_0^{\infty}dk\int_{-\infty}^{\infty}d\omega\; k^{\frac{d-2}{2}}\,
e^{-i\omega t + ik\theta}\, G_+(\omega,k)\,.
\end{equation}
In the second term of (\ref{eq:Gplus3}), $g_+(t,-\theta)$ stands for \ref{eq:gplusdef} with $e^{ik\theta}$ replaced by $e^{-ik\theta}$, while the prefactor, including the phase $e^{i\pi(d-2)/2 \cdot \lfloor \theta/\pi\rfloor}$, being evaluated at $\theta \in (0,\pi)$\cite{Jia:2026pmv}\footnote{If instead (\ref{eq:gplusdef}) is evaluated literally at $(-\theta)$, so that $\lfloor -\theta/\pi\rfloor=-1$, the result differs from $g_+(t,-\theta)$ as defined here by the factor $e^{-i\pi (d-2)/2}$. In terms of it, (\ref{eq:Gplus3}) reads $G_+(t,\theta)=g_+(t,\theta)+e^{i\pi(d-2)}[g_+(t,\theta)]_{\theta\rightarrow-\theta}$}.
The Wightman correlator and its Fourier transform in the eikonal regime were also discussed in \cite{Dodelson:2023nnr}, 
where a similar expression was rigorously derived\footnote{A rigorous analysis
in \cite{Dodelson:2023nnr} allowed for a more detailed formula of the Euclidean two-point
function, which includes summation over all the periodic images of the angular
variable $\theta$, i.e.,
\begin{equation}\label{eq:bulkconeformula}
G_E(\tau = it+\epsilon,\theta) = \sum_{n=0}^{\infty}
\Big( g_E(\tau,|\theta|+2\pi n) + e^{i\pi(d-2)}\, g_E(\tau, 2\pi-|\theta|+2\pi n)\Big)\,,
\end{equation}
with $g_E(\tau=it+\epsilon,\theta)$ given in the eikonal limit by an expression similar
to \eqref{eq:gplusdef}.}.

Henceforth, we focus on $g_+$, since the thermal two-point function can be completely
described in terms of $g_+$. Since $b\equiv\frac{\omega}{k}$ is finite in the eikonal
regime, it is convenient to change variables from $(\omega,k)$ to
$(b=\frac{\omega}{k},k)$, such that,
\begin{equation}\label{gplus0}
g_+(t,\theta) = \frac{e^{-i\pi(d-2)/4}\, e^{i\pi\frac{d-2}{2}\lfloor\frac{\theta}{\pi}\rfloor}}
{(2\pi)^{d/2+1}\,|\sin\theta|^{d/2-1}}
\int_0^{\infty}dk\int_{-\infty}^{\infty}db\; k^{\frac{d}{2}}\,
e^{-ik(b\,t-\theta)}\, G_+(b,k)\,.
\end{equation}

To proceed further we need to know $G_+(b,k)$. In the eikonal regime, this is roughly
proportional to the exponential of the phase shift acquired by a highly energetic particle
traversing the AdS black hole geometry. The bulk phase shift was computed to a few
leading orders in the temperature via CFT and holography in
\cite{Karlsson:2019dbd,Karlsson:2019txu}, 
where the two-dimensional case for a conical defect geometry was analysed from a
purely CFT perspective as well \cite{Kulaxizi:2019tkd}. 
In generic dimensions $d$ it was computed holographically in
\cite{Dodelson:2023nnr,Jia:2026pmv} by a systematic WKB computation. For further discussions on the
bulk phase shift, a partial list of references can be found here
\cite{Dodelson:2023nnr,Karlsson:2019txu,Kulaxizi:2019tkd,Kulaxizi:2018dxo,Karlsson:2019qfi,Meltzer:2019pyl,Karlsson:2020ghx,%
Giusto:2020mup,Bianchi:2020yzr,Parnachev:2020zbr,Chandorkar:2021viw,Ceplak:2021wak,Kim:2021hqy,Geytota:2021ycx,%
Bianchi:2022wku,deRham:2022hpx,Hartman:2022njz,Giusto:2023awo,Fardelli:2024heb,Salgarkar:2023sya,Chen:2024iuv}.

Based on the above, and in particular on \cite{Kulaxizi:2018dxo, Dodelson:2023nnr, Jia:2026pmv}, 
we consider,\footnote{
Here $(t,\theta)$ are the cylinder variables, and \eqref{eq:bulkphaseshift} is the
Regge limit of the unshifted Wightman function. The correlator continued to the Regge
sheet, as a function of the Regge variables $t'=t-\pi$, $\theta'=\theta-\pi$, is
related to it by $G_{\rm Regge}(t',\theta') = G_+(t'+\pi,\theta'+\pi)$; in momentum
space this shift is implemented by the pure AdS part, $k\pi(b-1)$, of the phase
$kS(b)$ \cite{Jia:2026pmv}.}
\begin{equation}\label{eq:bulkphaseshift}
G_+(b,k) = G^0_+(b,k) e^{ikS(b)}\,,
\end{equation}
where,
\begin{equation}\label{eq:Sdef}
S(b) = 2\int_{r_T}^{\infty}\frac{dr}{f(r)}\sqrt{b^2-\frac{f(r)}{r^2}}
= \int_0^{u_T}\frac{du}{f(u)}\sqrt{\frac{b^2-u f(u)}{u^3}}
\equiv b\,T(b) - \Theta(b)\,,
\end{equation}
with $(T(b),\Theta(b))$ equal to the time delay and angular deflection of a null
geodesic traversing the geometry,
\begin{equation}\label{eq:TThetadef}
\begin{aligned}
T(b) &= 2b\int_{r_T}^{\infty}\frac{dr}{f(r)}\frac{1}{\sqrt{b^2-\frac{f(r)}{r^2}}}
= b\int_0^{u_T}\frac{du}{f(u)}\frac{1}{\sqrt{u^3\left(b^2-uf(u)\right)}}\,,\\
\Theta(b) &= 2\int_{r_T}^{\infty}\frac{dr}{r^2}\frac{1}{\sqrt{b^2-\frac{f(r)}{r^2}}}
= \int_0^{u_T}du\,\frac{u}{\sqrt{u^3\left(b^2-uf(u)\right)}}\,,
\end{aligned}
\end{equation}
and $G^0_+(b,k)$ stands for the zero temperature correlator, in the standard
normalisation of the Wightman function (whose Fourier transform is the positive
spectral density), multiplied by the constant phase $e^{-i\pi(\Delta-\frac{d}{2}+1)}$
acquired in the Regge limit \cite{Jia:2026pmv},
\begin{equation}\label{eq:Gzero}
G^0_+(b,k) = e^{-i\pi\left(\Delta-\frac{d}{2}+1\right)}\,
\frac{2^{d+1-2\Delta}\pi^{1+\frac{d}{2}}}{\Gamma(\Delta) \Gamma\left(\Delta-\frac{d}{2}+1\right)}\,
k^{2\Delta-d}\left(b^2-1\right)^{\Delta-\frac{d}{2}}\,.
\end{equation}
In the conventions of \cite{Kulaxizi:2018dxo} the zero temperature correlator was normalised
with an additional factor $e^{i\pi\Delta}$, whereas the phase acquired by the correlator
on the Regge sheet is $e^{-i\pi\Delta}$ \cite{Jia:2026pmv}.
Note that by virtue of their definition in \eqref{eq:TThetadef}, the derivatives of
$T(b)$, $\Theta(b)$ satisfy,
\begin{equation}\label{eq:TThetaderrel}
b T'(b) - \Theta'(b) = 0\,.
\end{equation}

A null, real geodesic returning to the boundary exists for $b\in(1,b_c)$, and the turning point, $u_T$, is the positive solution of,
\begin{equation}\label{eq:turningpoint}
b^2 - u_T f(u_T) = 0 \ \overset{d=4}{\Longrightarrow} \ 
u_T = \frac{2(b^2-1)}{1+\sqrt{1-4\mu(b^2-1)}}\,.
\end{equation}
When $b=b_c$, the value of the turning point approaches that of the photon sphere and the particle gets trapped in a circular trajectory around the black hole,
\begin{equation}\label{eq:photonsphere}
\begin{aligned}
u_T &= u_{\rm photon} = \left(\frac{2}{d\mu}\right)^{\frac{2}{d-2}}\,,\\
b_c^2 &= 1 + \left(1-\frac{2}{d}\right)\left(\frac{2}{d\mu}\right)^{\frac{2}{d-2}}\,.
\end{aligned}
\end{equation}
In \cite{Dodelson:2023nnr}, the position space integral over the bulk phase shift was used to extract the so-called bulk cone singularities of the correlator, present in holographic
correlators when boundary points are connected by null geodesics in the bulk \cite{Hubeny:2006yu,Dodelson:2023nnr,Dodelson:2020lal}.

In \cite{Jia:2026pmv} the phase shift was analytically continued to the regime $b>b_c$. The purpose of the analysis was to investigate the physics in the regime of small impact parameters. Interestingly, a singularity pertaining to null geodesics bouncing off the black hole singularity was observed, via a similar computation, at $t=t_c+i\beta/2$, $\theta=0$. It lies outside the strip of analyticity of the Wightman function, while the two-sided correlator at $t=t_c-i\beta/2$ remains regular \cite{Jia:2026pmv}.
In this section, we will investigate the regime $0<b<1$, where the stress-tensor sector exhibits the spacelike singularities which correspond to imaginary null geodesics in the bulk.

From the point of view of the WKB approximation, in the region $0<b<1$, the WKB phase (a.k.a.\ the bulk phase shift) reduces to an exponential fall-off, representing the tunnelling of the particle through the region near the boundary where $f(r)/r^2>b^2$. To see this, notice that the term under the square root in \eqref{eq:Sdef} becomes negative. Thus, when the turning point is real and positive \footnote{here we think of the turning point in the WKB sense, as there exists no real null geodesic that returns to the boundary}, the phase in \eqref{eq:bulkphaseshift} reduces to an exponential. Evaluating the correlator in this case reveals no singularity.
It is possible, however, to consider the imaginary null geodesic discussed in the previous section. In this case, the turning point is also imaginary (corresponding to $u<0$) and the correlator remains a pure phase. The formulas in \eqref{eq:TThetadef} remain valid and real. This is because the real and positive turning point for $b>1$ becomes imaginary when $b<1$. \footnote{Alternatively, one can consider the expressions for the bulk time-delay and deflection obtained in $d=4$ in eqs (4.4) and (4.5) of \cite{Jia:2026pmv}.}.

Let us investigate the existence of a singularity in this case. Substituting into \eqref{gplus0} leads to,
\begin{equation}\label{eq:gplus1}
\begin{aligned}
g_+(t,\theta) &= \frac{e^{-i\pi\left(\Delta-(d-2)/4\right)}\,
e^{i\pi\frac{d-2}{2}\lfloor\frac{\theta}{\pi}\rfloor}}
{2^{2\Delta-d/2}\,|\sin\theta|^{d/2-1}\,\Gamma(\Delta) \Gamma\left(\Delta-\frac{d}{2}+1\right)}
\times\\
&\quad\times\int_0^{\infty}dk\int_0^{1}db \ k^{2\Delta-\frac{d}{2}}\,
(-1)^{\Delta-\frac{d}{2}}\left(1-b^2\right)^{\Delta-\frac{d}{2}}\,
e^{-ik\left(b\,(t-T(b))-(\theta-\Theta(b))\right)}\,.
\end{aligned}
\end{equation}
Integrating over $k$ results in,
\begin{equation}\label{eq:gplus1a}
\begin{aligned}
    g_+(t,\theta) & = \frac{e^{-i\pi\left(2\Delta-\frac{d}{2}+1\right)}\,
e^{i\pi\frac{d-2}{2}\lfloor\frac{\theta}{\pi}\rfloor}\,
\Gamma\left(2\Delta-\frac{d}{2}+1\right)}
{2^{2\Delta-d/2}\,|\sin\theta|^{d/2-1}\,\Gamma(\Delta)\,\Gamma\left(\Delta-\frac{d}{2}+1\right)} \times\\
& \quad \times \int_0^{1}\frac{(-1)^{\Delta-\frac{d}{2}}\left(1-b^2\right)^{\Delta-\frac{d}{2}}\,db}
{\left(b\,(t-T(b))-(\theta-\Theta(b))-i\epsilon\right)^{2\Delta-\frac{d}{2}+1}}\,,
\end{aligned}
\end{equation}
where the $i \epsilon$ comes from requiring convergence of the integral over $k$.
We have a candidate singularity when,
\begin{equation}\label{eq:sing_conditions}
\begin{aligned}
&b\,t - \theta - b\,T(b) + \Theta(b) = 0\,,\\
&\frac{\partial}{\partial b}\left(b\,t-\theta-b\,T(b)+\Theta(b)\right) = 0\,,
\end{aligned}
\end{equation}
which leads to the pinch condition,
\begin{equation}\label{eq:sing_position}
\begin{aligned}
t &= T(b)\,,\\
\theta &= \Theta(b)\,,
\end{aligned}
\end{equation}
where we also used the derivative relations \eqref{eq:TThetaderrel}. The functional
form of the correlator near the singularity can be computed by expanding around the
pinch singularity. This is done by expressing $b$ as $b=b(t)+\delta b$, where $b(t)$
satisfies $T(b(t))=t$. A second-order expansion then yields,
\begin{equation}\label{eq:TThetaTaylor}
-\theta + \Theta(b) + b\,(t-T(b)) \approx -\theta + \Theta(b(t))
- \frac{1}{2}T'(b(t))\,\delta b^2\,,
\end{equation}
where we have used \eqref{eq:TThetaderrel} to eliminate the second derivatives of
$T(b)$, $\Theta(b)$.

Near the singularity the integral becomes,
\begin{equation}\label{eq:gplus2}
\begin{aligned}
g_+(t,\theta) &\simeq \frac{e^{-i\pi\left(2\Delta-\frac{d}{2}+1\right)}\,
e^{i\pi\frac{d-2}{2}\lfloor\frac{\theta}{\pi}\rfloor}\,
\Gamma\left(2\Delta-\frac{d}{2}+1\right)}
{2^{2\Delta-d/2}\,|\sin\theta|^{d/2-1}\,\Gamma(\Delta)\,\Gamma\left(\Delta-\frac{d}{2}+1\right)} \times\\
&\quad \times (-1)^{\Delta-\frac{d}{2}}\left(1-b^2\right)^{\Delta-\frac{d}{2}} \int_{-\infty} ^{\infty}\frac{d\delta b} {\left(-\theta+\Theta(b(t))-\frac{1}{2}T'(b(t)) \delta b^2\right)^{2\Delta-\frac{d}{2}+1}}\\
&= \frac{e^{-i\pi\left(2\Delta-\frac{d}{2}+1\right)}\,(-1)^{\Delta-\frac{d}{2}}\, e^{i\pi\frac{d-2}{2}\lfloor\frac{\theta}{\pi}\rfloor}\, \Gamma\left(2\Delta-\frac{d}{2}+\frac{1}{2}\right)} {2^{2\Delta-d/2}\,|\sin\theta|^{d/2-1}\,\Gamma(\Delta)\,\Gamma\left(\Delta-\frac{d}{2}+1\right)} \times \\
& \quad \times  \sqrt{\frac{-2\pi}{T'(b(t))}} \frac{\left(1-b(t)^2\right)^{\Delta-\frac{d}{2}}} {\left(-\theta+\Theta(b(t))\right)^{2\Delta-d/2+1/2}}\,.
\end{aligned}
\end{equation}
Close to the singularity, the expression $-\theta+\Theta(b(t)) = b(t)\left(t-t_{BC}(\theta)\right)$
holds, where $t_{BC}(\theta)\equiv T(b(t))\big|_{\Theta(b(t))=\theta}$. Substituting $g_+(t,\theta)$ back into \eqref{eq:Gplus3} yields,
\begin{equation}\label{eq:Gplussingularity}
\begin{aligned}
G_+(t,\theta) &\approx \frac{e^{-i\pi\left(2\Delta-\frac{d}{2}+1\right)}\,(-1)^{\Delta-\frac{d}{2}}\,
e^{i\pi\frac{d-2}{2}\lfloor\frac{\theta}{\pi}\rfloor}\,
\Gamma\left(2\Delta-\frac{d}{2}+\frac{1}{2}\right)}
{2^{2\Delta-d/2}\,|\sin\theta|^{d/2-1}\,\Gamma(\Delta)\,\Gamma\left(\Delta-\frac{d}{2}+1\right)}
\sqrt{\frac{2\pi}{-T'(b(t))}}\;
\frac{\left(1-b(t)^2\right)^{\Delta-\frac{d}{2}}}{b(t)^{2\Delta-d/2+1/2}}
\;\times\\
&\quad\times\left(\frac{1}{\left(t-t_{BC}(\theta)\right)^{2\Delta-d/2+1/2}}
+ e^{i\pi(d-2)/2}\,\frac{1}{\left(t-t_{BC}(-\theta)\right)^{2\Delta-d/2+1/2}}\right).
\end{aligned}
\end{equation}
Hence, it appears as though the Wightman correlator in the eikonal regime, i.e.\ its Regge-limit part obtained by Fourier transforming \eqref{eq:bulkphaseshift} \cite{Jia:2026pmv}, exhibits a singularity related to the trajectory of a null, imaginary geodesic in the AdS black hole geometry.

We now return to the black brane case to compare with the results obtained earlier. Scaling the horizon over the AdS length to infinity, reduces \eqref{eq:adssch} to,
\begin{equation}\label{eq:BBmetric}
ds^2 = -\frac{1-u^{d/2}}{u}\,dt^2 + \frac{1}{u}\,dx^2
+ \frac{1}{4u^2\left(1-u^{d/2}\right)}\,du^2\,,
\end{equation}
which is equal to \eqref{BBAdS5} in $d=4$. We need not perform the computations anew, but rather set $\theta\approx x$ in \eqref{eq:Gplussingularity}. Since we are considering the large radius limit, where $\theta\ll1$, we have $\sin\theta\approx\theta\approx x$ and $\lfloor\theta/\pi\rfloor\approx0$. Hence, the thermal two point function near the singularity will behave as follows,
\begin{equation}\label{eq:octp6}
\begin{aligned}
G_+(t,x) &\approx \frac{e^{-i\pi\left(2\Delta-\frac{d}{2}+1\right)}\,(-1)^{\Delta-\frac{d}{2}}\, \Gamma\left(2\Delta-\frac{d}{2}+\frac{1}{2}\right)} {2^{2\Delta-d/2}\,|x|^{d/2-1}\,\Gamma(\Delta)\,\Gamma\left(\Delta-\frac{d}{2}+1\right)} \sqrt{\frac{2\pi}{-T'(b(t))}} \frac{\left(1-b(t)^2\right)^{\Delta-\frac{d}{2}}}{b(t)^{2\Delta-d/2+1/2}}
\times\\
&\quad\times\left(\frac{1}{\left(t-t_{BC}(x)\right)^{2\Delta-d/2+1/2}}
+ e^{i\pi(d-2)/2}\,\frac{1}{\left(t-t_{BC}(-x)\right)^{2\Delta-d/2+1/2}}\right),
\end{aligned}
\end{equation}
where $t_{BC}(x)$ now satisfies,
\begin{equation}\label{eq:XTrelations}
-x + X(b(t)) = b(t)\left(t-t_{BC}(x)\right),
\end{equation}
and $X$ and $T$ are given by \eqref{eq:TThetadef} with
$f(u)=\frac{1}{u}-u^{d/2-1}$. They reduce to \eqref{eq:null_geo_t_x_def} in $d=4$. 

It is useful to check that the prefactor of the singular terms in \eqref{eq:octp6} is well-defined. In particular, note that $T'(b)$ is neither divergent, nor vanishing for $0<b<1$, and is continuous at $b=0$.
\begin{equation}\label{eq:Tprimedef}
T'(b) = \frac{2(2\pi)^{3/2}}{\Gamma\!\left(-\frac{1}{4}\right)^2}\,
\frac{1}{\left(1-b^2\right)^{5/4}} > 0\,.
\end{equation}
The branch of the square root $\sqrt{2\pi/(-T'(b(t)))}$ in \eqref{eq:octp6} is fixed by the $i\epsilon$ prescription in \eqref{eq:gplus1a}. 

\section{Discussion}\label{sec:discussion} 

In this paper, we investigated the stress-tensor sector of the thermal two point function of generic conformal dimension $\Delta$ in holographic CFTs. Using OPE resummation we found that it exhibits singularities when the operators are situated a spacelike distance apart on the boundary, connected in the bulk by imaginary geodesics. They are similar in nature to the standard bulk-cone singularities. We then studied the position space Wightman two-point in the Regge limit\footnote{Note that this kinematic regime includes an analytic continuation that takes the Wightman function on the second sheet, see {\it e.g} \cite{Kulaxizi:2018dxo}. }, by Fourier transforming the bulk phase shift. Evaluating the bulk phase shift over imaginary geodesics with $\omega<k$, revealed singularities in the corresponding correlator precisely at the positions of the bulk imaginary null geodesics. 


The results are suggestive, but the question of whether the full thermal, holographic correlator (as opposed to its stress-tensor sector) exhibits bulk-cone type singularities related to the existence of null, complex geodesics in the bulk, has not been completely answered.
In particular, it would be interesting to consider the contribution of double-trace operators to the correlator, along the lines of \cite{Buric:2025anb, Buric:2025fye}
(see also \cite{Barrat:2025nvu,Niarchos:2025cdg,Barrat:2025twb}).



Avenues for further investigations consist of different holographic backgrounds, such AdS-RN -- recently investigated in \cite{Ceplak:2025dds, Dodelson:2025jff, AliAhmad:2026wem} -- or AdS-Kerr black holes. A special case of interest are also the Robinson-Trautman spacetimes \cite{BernardideFreitas:2014eoi, Skenderis:2017dnh}, recently investigated in the holographic context in \cite{Castro:2025itb}.

In complete analogy, but perhaps with a different perspective, one can consider asymptotically flat or de Sitter geometries. Retarded thermal correlators enter in certain computations pertaining to gravitational waves, e.g. \cite{Casals:2009zh,Casals:2013mpa, Wardell:2014kea}, and it will be interesting to see if the observations herein have an imprint there. Regarding de Sitter spacetime, it is natural to ask how the complex geodesics and similar geometric objects discussed in \cite{Chapman:2022mqd, Aalsma:2022eru,Heller:2025kvp,Bohra:2025mhb} are related to this work.

\bigskip

\section*{Acknowledgments}
We thank I. Buric, N. Ceplak, I. Gusev, H. Liu, S. Valach for useful discussions.
I.J.A. acknowledges funding by ANID FONDECYT grants 1261016 and 1231133.
CE  was supported in part by the Irish Research Council Government of Ireland Postgraduate Fellowship  under project award number GOIPG/2022/887.
The work of YJ is funded in part by the China Scholarship
Council (project ID 202309120013). MK acknowledges support in part from COST Action CA22113, for a short scientific mission to NKUA, Greece, where part of this work was completed. 
MK is thankful to the Physics Department of NKUA, Greece, for hospitality.
The work of AP is supported in part by Taighde \'Eireann - Research Ireland under  grant agreements 22/EPSRC/3832 and SFI-22/FFP-P/11444.
IJA, MK and AP thank the Simons center for Geometry and Physics for hospitality and support during the program ’Black Hole Physics from Strongly Coupled Thermal Dynamics’, where part of this work was completed.


\appendix
\section{Details of Boundary Ansatz}
\label{app_bndry_anz}

In this section we describe the process of computing the boundary anstaz coefficients $a^i_{j, k}$ in \ref{eq:corr_nonstand_gen}. We show how to find the differential operator $\mathcal{D}$ in (\ref{eq:diff_op_psi}) and the recursion relation for $a^i_{j, k}$.  

We begin with the equation of motion for a minimally coupled scalar field $\phi$ as in (\ref{eq:KGEq}),
\begin{equation}
   \left( \Box- \Delta(\Delta-4) \right)\phi(r,t, \vec{x}) = 0\,.
   \label{eq:scalar_bulk_eq}
\end{equation}
In the black brane background (\ref{eq:bcknd}), (\ref{eq:scalar_bulk_eq}) is given explicitly as follows,
\begin{equation}
    \left(r^2 f(r) \partial_r^2 + \left(r^2 f'(r)+5 r f(r) \right)\partial_r  +\frac{1}{r^2 f(r)}\partial_{\tau}^2 +\frac{1}{r^2} \sum_{i=1}^3 \partial_{x_i}^2-\Delta(\Delta-4)\right) \phi(r, \tau, \vec{x}) = 0\,,
    \label{eq:scalar_bulk_eq_explicit}
\end{equation}
with,
\begin{equation}
     f(r) = 1-\frac{1}{r^4}\,.
     \label{eq:metric_f_black_brane}
\end{equation}
We convert the cartesian spatial coordinates $\vec{x} = (x_1, x_2, x_3)$ into polar coordinates $(|\vec{x}|, \theta_1, \theta_2)$ and use spherical symmetry such that $\phi(r, \tau, |\vec{x}|, \theta_1, \theta_2)= \phi(r, \tau, |\vec{x}|)$. The resulting equation is given by,
\begin{equation}
      \left(r^2 f(r) \partial_r^2 + \left(r^2 f'(r)+5 r f(r) \right)\partial_r  +\frac{1}{r^2 f(r)}\partial_\tau^2 +\frac{1}{ r^2}\partial_{|\vec{x}|}^2 + \frac{2}{|\vec{x}|r^2}\partial_{|\vec{x}|}  -\Delta(\Delta-4)\right) \phi(r, \tau, |\vec{x}|) = 0\,.
      \label{eq:scalar_bulk_eq_polar}
\end{equation}
\
We introduce new coordinates,
\begin{equation}
    \rho = r \tau\,, \quad \lambda^2 = 1+ r^2 \left(|\vec{x}|^2+\tau^2 \right)\,,
    \label{eq:def_rholabda_app}
\end{equation}
and factor out the pure AdS solution,
\begin{equation}
    \phi(r,\rho, \lambda) = \left( \frac{r}{\lambda^2}\right)^\Delta\psi(r, \rho, \lambda)\,.
    \label{eq:def_pasi}
\end{equation}
The resulting equation is given by,
\begin{equation}
\begin{aligned}
     \big(\partial_r^2 + C_1 \partial^2_\lambda  + C_2 \partial^2_{\rho} + & C_3 \partial_r \partial_\lambda +  C_4 \partial_r \partial_{\rho}  +C_5 \partial_\lambda \partial_{\rho} 
       \\
      & + C_6 \partial_r + C_7 \partial_\lambda + C_8 \partial_{\rho}  + C_9 \big)\psi(r, \rho , \lambda)=0\\,
    \label{eq:eom_psi_nonstand}
\end{aligned}
\end{equation}
where the coefficients are given as follows,
\begin{equation}
    \begin{aligned}
        C_1 = & \frac{\left(\lambda^2-1\right)^2 f(r)^2 + f(r) \left( - \rho^2 + \lambda^2 - 1\right) + \rho^2}{r^2 \lambda^2 f(r)^2}\\
      C_2 = & \frac{\rho^2 f(r)^2 + 1}{r^2 f(r)^2}\\
         C_3 = &\frac{2\left(\lambda^2-1 \right)}{r \lambda}\\
        C_4 = & \frac{2 \rho}{r}\\ 
        C_5 = &  \frac{2 \rho}{r^2 \lambda f(r)^2}\left(1+f(r)^2\left( \lambda^2-1\right) \right)\\
        C_6 = & \frac{f'(r)}{f(r)} + \frac{4 \Delta + \lambda^2 \left(5-2 \Delta \right)}{r \lambda^2} \\ 
        C_7 =&\frac{1}{r^2 \lambda^3 f(r)^2} \Big( -\left( 1+4 \Delta\right)\rho^2 + \lambda^2 - f(r)^2 \left( \lambda^2-1\right)\left( 2 \Delta \left( \lambda^2-2\right) - 1-5\lambda^2\right)  \\
        & \quad -f(r) \left( -\rho^2 - 2 \lambda^2 - 1 + 4 \Delta \left( -\rho^2+\lambda^2-1\right) - f'(r) r \lambda^2 \left(\lambda^2-1 \right)\right)\Big) \\
        C_8 =& \frac{\rho}{r^2 \lambda^2 f(r)^2}\left( f(r)^2\left( 5 \lambda^2 - 2 \Delta \left( \lambda^2-2\right)\right) + f'(r)f(r)r \lambda^2 - 4 \Delta\right) \\ 
        C_9 =& \frac{\Delta}{r^2 \lambda^4 f(r)^2} \Big( f(r)^2 \left(4 + 4 \lambda^2 - 4 \lambda^4 + \Delta \left( \lambda^2-2\right)^2 \right)  - 2\left(-2\left(1+\Delta \right)\rho^2 + \lambda^2 \right) \\
        &\quad + f(r) \left( -4 \rho^2 - 4 - 2 \lambda^2+4\lambda^4 + \Delta\left(-4\rho^2-\left( \lambda^2-2\right)^2 \right) - f'(r)  r \lambda^2 \left( \lambda^2-2\right)\right)\Big)
        \label{eq:eom_coeff_nonstand}
    \end{aligned}
\end{equation}

The central part of the boundary anstaz method of \cite{Fitzpatrick:2019zqz} is the assertion that (\ref{eq:eom_psi_nonstand}) can be solved order-by-order for large $r$, and that $\psi(r, \rho, \lambda)$ is given by a finite number of polynomial terms at each order in $r$, such that a suitable ansatz for $\psi(r, \rho, \lambda)$ is given by,
\begin{equation}
     \psi(r, \rho, \lambda)  = \sum_{i=0}^\infty \sum_{j=0}^i \sum_{k=-i}^{2i-j} \frac{\rho^{2j} \lambda^{2k}}{r^{4i}} (-1)^j a^i_{j, k}\,, \quad a^0_{0, 0}=1\,.
     \label{eq:psi_anz_app}
\end{equation}
Notice that this is a series expansion that includes negative powers in $\lambda$. This is not an issue in Euclidean signature since $\lambda>1$.

To arrive at (\ref{eq:psi_anz_app}), we have imposed normalisability, ie. $\phi(r, \tau, |\vec{x}|) \sim r^{-\Delta}\,, \ r \gg 1$, and regularity at $\lambda = 1$. The regularity condition relies on the fact that the point $|\vec{x}| = \tau = 0$ is not a special bulk point, {\it i.e.} $\phi(r, \tau, |\vec{x}|)$ does not experience an singularity at this point.

The boundary ansatz method \cite{Fitzpatrick:2019zqz} is developed in Euclidean signature. It is unclear what the correct set of conditions is in Lorentzian signature. Here, we restrict ourselves to spacelike boundary separation, such that the regularity condition can be readily extended to Lorentzian signature.

One can compute a recursion relation between various $a^i_{j, k}$ by substituting the ansatz (\ref{eq:psi_anz_app}) into the equation of motion (\ref{eq:eom_psi_nonstand}) and performing index redefinition such that $(r, \rho, \lambda)$ appear as overall factors only. We ignore the fact that the summations in $(i, j, k)$ are bounded, and instead explicitly set all coefficients $a^i_{j, k}$ with values of $(i, j, k)$ outside the summation bounds to zero. 

The recursion relation used for computing $a^i_{j, k}$ is as follows,
\begin{equation}
    \begin{aligned}
        a^i_{j, k} = \frac{1}{4(4i-k)(k-\Delta)}\Big( & 4(1+2i-j-k)(1-2i+j+k-\Delta)a^i_{j, k-1} \\
        + & (2+6j+4j^2) a^i_{j+1, k-1} \\
        - & (6-4i+2j+2k-\Delta)^2 a^{i-2}_{j, k-1}\\
        + & 4(k-\Delta)(7-4i+2j+2k-\Delta) a^{i-2}_{j, k}\\
        - & 4(k-\Delta)(1+k-\Delta)a^{i-2}_{j, k+1}\\
        + & 4(k-\Delta)(1+k-\Delta)a^{i-1}_{j-1, k+1}\\
        + & (32i^2 + 8(1+j+k)(2+j+k-\Delta))a^{i-1}_{j, k-1} \\
        - & (16i (3+2j+2k-\Delta) - \Delta^2)a^{i-1}_{j, k-1}\\
        - & 2(k-\Delta)(15-16i + 8j+6k-2\Delta)a^{i-1}_{j, k}\\
        + & 4(k-\Delta)(1+k-\Delta)a^{i-1}_{j, k+1}\Big)\,.
        \label{eq:recursion_nonstand}
    \end{aligned}
\end{equation}

To compute the coefficients $a^i_{j, k}$ via (\ref{eq:recursion_nonstand}), we follow the procedure described in \cite{Buric:2025fye}, which vastly optimises the computation. 

\subsection{Holographic OPE from CFT}
In this section we provide a one to one correspondence between the standard CFT OPE expansion and the series expansion given in (\ref{eq:corr_nonstand_gen}). 
We are interested in computing the thermal scalar correlator,
\begin{equation}
    \langle \phi(\tau, \vec{x}) \phi(0,0) \rangle_{\beta}\,,
    \label{eq:two_point_function}
\end{equation}
in a strongly coupled holographic CFT. Here, $\phi$ is a scalar primary with conformal dimension $\Delta$.
In holographic theories the correlator can be decomposed as follows,
\begin{equation}
    \langle \mathcal{O}(\tau, \vec{x}) \mathcal{O}(0,0)\rangle_\beta = G(\tau, \vec{x}) = G_T(\tau, \vec{x}) + G_{[\mathcal{O} \mathcal{O}]}(\tau, \vec{x})\,.
    \label{eq:def_G_app}
\end{equation}
Here, $G_{[\mathcal{O} \mathcal{O}]}(\tau, \vec{x})$ denotes the sector composed of double-trace contributions. $G_T(\tau,\vec{x})$ denotes the stress-tensor sector of the correlator which is determined by the multi-stress tensor contributions, $T^n$. 
These operators are conformal primaries built from products of $n$ stress-tensors. In general dimensions, $T^n$ have conformal dimension $\Delta_T = n d$ to leading order in large central charge $C_T$. By contracting indices, the spin $J$ can range from $[0, 2n]$ for $n>1$. \footnote{The stress-tensor operator $T_{\mu \nu}$ is conserved and traceless with spin $J=2$.} Notice we have not included operators with derivatives since we are interested in correlators in flat space.


We are interested in computing $G_T$ in the large volume limit, which can 
equivalently be computed by the conformal block decomposition of the heavy-heavy-light-light, given that stress tensor sector thermalises (see \cite{Karlsson:2021duj})
\begin{equation}
    \langle \phi(\tau, \vec{x}) \phi(0,0) \rangle_{\beta} \Big|_{T^n} = \langle \mathcal{O}_H |\phi(\tau, \vec{x}) \phi(0,0)  | \mathcal{O}_H \rangle_{T^n}
    \label{eq:2pt_corr}
\end{equation}
We begin with the heavy-heavy-light-light (HHLL) correlator on $\mathbb{R}^4$, which can be expanded using conformal partial waves (CPW) as follows \cite{Dolan:2000ut,Dolan:2003hv},
\begin{equation}
    \langle \mathcal{O}_H(0)  \phi(z, \zb) \phi(1) \mathcal{O}_H(\infty)\rangle \Big|_{T^n} = \frac{1}{[(1-z)(1-\zb)]^\Delta} \sum_{\Delta_T, J} c_\beta(\Delta_T, J) W_{\Delta_T, J}(1-z, 1-\zb)\,. 
    \label{eq:4pt_corr}
\end{equation}
The coefficients $c_\beta(\Delta_T, J)$ denote the product of OPE coefficients $\lambda_{\mathcal{O}\mathcal{O}T^n}$ with the thermal one-point functions $\langle T^n \rangle_{\beta}$. The conformal partial waves $W_{\Delta_T, J}(z, \zb)$ are given by,
\begin{equation}
    W_{\Delta_T, J}(z, \zb) = \frac{z\zb}{z-\zb} \left( k_{\Delta_T + J}(z)k_{\Delta_T - J - 2}(\zb) - (z \leftrightarrow \zb) \right)\,,
\label{eq:conf_partial_four_expression}
\end{equation}
where $k_{\a}(x) = x^{\a/2} {}_2F_1\left(\frac{\a}{2}, \frac{\a}{2}, \a, x \right)$. We map to $\mathbb{R} \times S_R^{d-1}$ via the following transformation,
\begin{equation}
z = e^{-\frac{\tau}{R} - i \frac{|\vec{x}|}{R}}\,,\  \zb = e^{-\frac{\tau}{R} + i \frac{|\vec{x}|}{R}}\,,
\label{eq:map_plane_R_cross_S_three}
\end{equation}
where the CPW transform as follows,
\begin{equation}
    W_{\Delta_T, J}(z, \zb)\Big|_{\mathbb{R}\times S^3_R} = R^{-2\Delta} [(1-z)(1-\zb)]^{\Delta/2}  W_{\Delta_T, J}(z, \zb) \Big|_{\mathbb{R}^4}\,.
    \label{eq:conf_partial_S_three}
\end{equation}

By the operator-state correspondence, the HHLL correlator (\ref{eq:4pt_corr}) on $\mathbb{R} \times S_R^{3}$ can be seen as (\ref{eq:2pt_corr}) but on $\mathbb{R} \times S_R^{3}$. In this paper we focus on the infinite volume limit $R \to \infty$, which leads to,
\begin{equation}
    G_T(Z, \Zb) = \frac{1}{(Z \Zb)^{\Delta}} \sum_{\Delta_T, J} c_\beta(\Delta_T, J) \ (Z \Zb)^{\frac{\Delta_T-J}{2}} \frac{Z^{J+1}-\Zb^{J+1}}{Z-\Zb}\,,
    \label{eq:CFT_corr_OPE}
\end{equation}
where,
\begin{equation}
    Z \approx (\tau + i x)\,, \quad\overline{Z} \approx (\tau - i x)\,.
    \label{eq:zzbar_planar}
\end{equation}

We would like to compare this CFT series expansion with the holographic computation (\ref{eq:corr_nonstand_gen}).
Given (\ref{eq:zzbar_planar}), the $AdS_5$ coordinates used in section (\ref{sec:boundry_ansatz}) are given by,
\begin{equation}
    \lambda^2 = 1 + r^2 (Z \overline{Z})^2\,,  \ \rho^2 = \frac{r^2}{4} \left(Z + \overline{Z}\right)^2\,,
    \label{eq:lambda_rho_Z_Zbar}
\end{equation}
which allows us to rewrite \ref{eq:corr_nonstand_gen} as follows,
\begin{equation}
    G_T(z, \zb) = \frac{1}{(Z \Zb)^{\Delta}} \sum_{i=0}^\infty \sum_{j=0}^\infty \left(-\frac{1}{4}\right)^j a^i_{j, 2i-j} (Z \Zb)^{2i-k} (Z - \Zb)^{2j}\,.
    \label{eq:hol_corr_OPE}
\end{equation}

Comparing (\ref{eq:CFT_corr_OPE}) to (\ref{eq:corr_nonstand_gen}), we obtain a relationship between the $a^i_{j, k}$ and the OPE coefficients $c_\beta$,
\begin{equation}
\begin{aligned}
       \sum_{J=0, J \in 2\mathbb{N}}^{2n} c_\beta(4n, J) &  \left(Z \Zb\right)^{2n-J/2} \frac{Z^{J+1}-\Zb^{J+1}}{Z-\Zb} \\
      & = \sum_{j=0}^n a^n_{j, 2n-j} \left(-\frac{1}{4}\right)^j\left(Z \Zb \right)^{2n-j} (Z- \Zb)^{2j}\,, \ \forall Z, \Zb\,,
\end{aligned}
    \label{eq:rel_hol_CFT_series}
\end{equation}
from which we can identify the OPE coefficients as a function of the boundary ansatz coefficients,
\begin{equation}
    c_{\beta}^{4n, 2s} = \sum_{j=s}^n a^n_{j, 2n-j}\frac{(-1)^s}{4^j} \binom{2j+1}{j-s}\,,
\end{equation}
and the also inverse relation,
\begin{equation}
    a^n_{j, 2n-j} = \sum_{s=j}^{n} c_{\beta}^{4n, 2s}\frac{(-4)^j (s+j)!}{(2j+1)!(s-j)!} (2s+1)\,.
\end{equation}

Note that we are able to directly associate the contribution of $[T^n]$ to the OPE with the term of order $r^{-4n}$ in the boundary ansatz (\ref{eq:psi_anz_app}). Upon reinstating the dependence of temperature $\beta$, it is clear that both expansions are expansions in temperature. 

Writing the expressions of series coefficients that control the singularities at finite $t$, (\ref{eq:corr_sing_t0}), as a function of OPE coefficients it becomes clear that the relevant asymptotic structure arises from a nontrivial weighted sum over coefficients with varying scaling dimension and spin. 
\begin{equation}
    \Lambda_n(t) = \sum_{i=\lceil \frac{n}{2}\rceil}^n \sum_{s=2i-n}^i t^{4i} \frac{(-4)^{2i-n}(2s+1)(s+2i-n)!}{(4i-2n+1)!(s-2i+n)!} c_{\beta}^{4i, 2s}\,.
\end{equation}

\section{Singularities for small $\xx$}
Given that the singularities at fixed real time are bounded from below, it is natural to ask whether there exist singularities that are instead bounded above by this same value. If such singularities exist we may expect them to correspond to complex values of $t$, which cannot be sampled numerically.
Instead we would like to keep $|\vec{x}|$ fixed, such that we are sampling the finite region $|\vec{x}| \in (0, |\vec{x}|_c)$, where $|\vec{x}_c|$ is given by (\ref{eq:Xforbzero}).

The form of the correlator in (\ref{eq:corr_nonstand_gen}) requires some additional manipulation in order to bring it into the form of a power series where $\xx$ remains fixed. 
We first rewrite \eqref{eq:corr_nonstand_gen} as,
\begin{equation}
\begin{aligned}
    G_T(t, \xx) &= \frac{1}{(\xx^2-t^2)^{\Delta}}\sum_{i=0}^\infty \sum_{j=0}^i \xx^{4i} \left( \frac{1}{1-\frac{t^2}{\xx^2}} - 1\right)^{j} \left(1-\frac{t^2}{\xx^2}\right)^{2i}a^i_{j, 2i-j}\\
    &=\frac{1}{(\xx^2-t^2)^{\Delta}}\sum_{i=0}^\infty \sum_{j=0}^i \sum_{k=0}^j  \xx^{4i} \binom{j}{k}(-1)^{j-k} \left(1-\frac{t^2}{\xx^2}\right)^{2i-k}a^i_{j, 2i-j}\,,
\end{aligned}
\end{equation}
where the factors of $t$ are absorbed into $(\xx^2-t^2)$ by a careful application of the binomial expansion.

We can exchange the summation intervals of $j, k$ to obtain the following series expansion, 
\begin{equation}
    \begin{aligned}
        G_T(t, \xx) = \frac{1}{(\xx^2-t^2)^{\Delta}} \sum_{i=0}^\infty \sum_{k=0}^i   \xx^{4i}  \left(1-\frac{t^2}{\xx^2}\right)^{2i-k} \sum_{j=k}^i a^i_{j, 2i-j} \binom{j}{k}(-1)^{j-k}\,,
    \end{aligned}
\end{equation}
which, via the same steps as in (\ref{eq:corr_lc_seris}), can be brought into a power series about the lightcone,
\begin{equation}
    G_T(t, |\vec{x}|) = \frac{1}{(\xx^2-t^2)^{\Delta}} \sum_{n=0}^\infty \left(1-\frac{t^2}{|\vec{x}|^2}\right)^n \tilde{\Lambda}_n(|\vec{x}|)\,,
    \label{eq:small_x_pwr_series}
\end{equation}
with series coefficients,
\begin{equation}
    \tilde{\Lambda}_n(|\vec{x}|) = \sum_{k=\lceil n/2 \rceil}^n \sum_{s=2k-n}^k |\vec{x}|^{4k} (-1)^{s-2k+n}\binom{s}{2k-n} a^k_{s, 2k-s}\,.
\end{equation}

In contrast to the series coefficients at fixed $t$ (\ref{eq:corr_sing_t0}), the coefficients $\tilde{\Lambda}_n(\xx)$ approach the following asymptotic form,
\begin{equation}
    \lim_{n \to \infty}  \Lambda_n = \frac{|A(\Delta, \xx)|}{|B(\xx)|^n}n^{2\Delta-b(\Delta, \xx)} \cos{\left(n \arg B(\xx) + \arg A(\Delta, \xx) \right)}\,.
    \label{eq:small_x_asymp}
\end{equation}
Upon substituting (\ref{eq:small_x_asymp}) into (\ref{eq:small_x_pwr_series}) and replacing summation with integration we find,
\begin{equation}
\begin{aligned}
     G_T(t, \xx) & \approx \frac{\Gamma\left(2\Delta - b(\Delta, \xx)+1\right)}{2 \left(\xx^2-t^2 \right)^{\Delta}} 
 \Bigg
 ( \left(- A(\Delta, \xx)\log\left( \frac{\xx^2-t^2}{\xx^2 B(\xx)}\right)\right)^{b(\Delta, \xx)-2\Delta-1}\\
   & \qquad + \left(- A^*(\Delta, \xx)\log\left( \frac{\xx^2-t^2}{\xx^2 B^*(\xx)}\right)\right)\Bigg)^{b(\Delta, \xx)-2\Delta-1}
   \label{eq:small_x_near_sing}
\end{aligned}
\end{equation}

The two terms in (\ref{eq:small_x_near_sing}) produce two separate cases where (\ref{eq:small_x_pwr_series}) no longer converges. For generic complex $B(\xx)$, there exists a singularity at generic complex $t$ at,
\begin{equation}
      t = \xx \sqrt{1-B(\xx)}\,,
\end{equation}
in addition to the images given by $(-t, t^*, -t^*)$. The correlator near such a singularity behaves as,
\begin{equation}
\begin{aligned}
     G_T(t, \xx) \approx & - \frac{A(\Delta, \xx)\Gamma(2\Delta-b(\Delta, \xx)+1)}{(-2)^{2\Delta -b(\Delta, \xx)+2} \xx^{1-b(\Delta, \xx)}} \\
    & \frac{B(\xx)^{\Delta - b(\Delta, \xx)-1}}{(1-B(\xx))^{\Delta - b/2 + 1/2}} \left( \frac{1}{t - \xx \sqrt{1-B(\xx)}}\right)^{2\Delta-b(\Delta, \xx)+1}
\end{aligned}
\end{equation}
The strength of the singularity, dictated by the function $b(\Delta, \xx)$, is no longer well approximated by a constant. One can view this as a consequence of interpolating between the case of $\xx =0$, where $b(\Delta, 0) = 3$, and $\xx > \xx_c$ with $b(\Delta, \xx) = 5/2$. A required property of $b(\Delta, \xx)$ is that it grows more slowly than $\Delta$ in the geodesic limit of $\Delta \gg 1$.

Fitting coefficients to the form (\ref{eq:small_x_asymp}) is more difficult due to the 5 unknown parameters it contains. The oscillatory nature also prevents the use of common methods like taking ratios of coefficients, as they become unbounded. 

\begin{figure}
    \centering
    \includegraphics[width=0.45\linewidth]{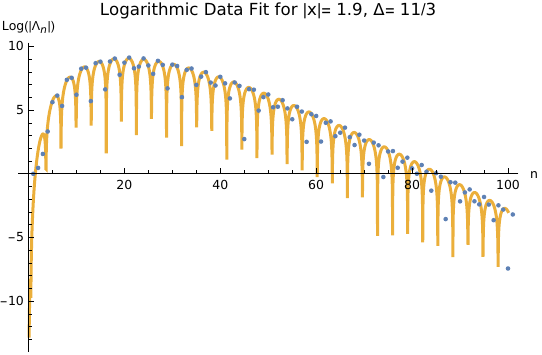}
    \includegraphics[width=0.45\linewidth]{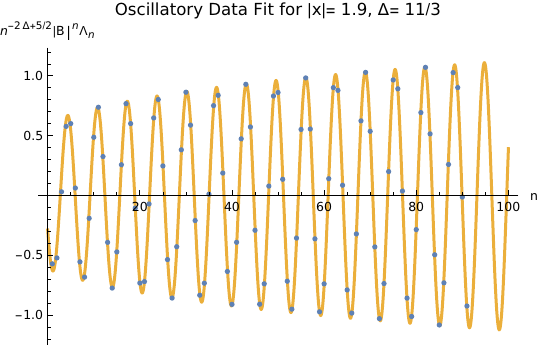}
    \caption{Data and fitted function for $\log(|\Lambda_n|)$(left) and $n^{-2\Delta + 5/2}|B|^n \Lambda_n$ (right) for $\Delta= 11/3$. While the logarithmic fit captures the amplitude envelope, it struggles to fit the oscillations.}
    \label{fig:log_osc_fit_compare}
\end{figure}

Instead we take advantage of the separation of scales governing the amplitude envelope and the oscillations. First, we perform a nonlinear fit on $\log{\tilde{\Lambda}(\xx)}$ using the full asymptotic form (\ref{eq:small_x_asymp}). The parameters governing the amplitude envelope, $|A(\Delta, \xx)|, |B(\xx)|, b(\Delta, \xx)$ are well approximated in this case, while the estimation of the oscillatory parameters $(\arg{A(\Delta, \xx)}, \arg{B(\xx)})$ is invalid. We disregard these oscillatory estimates and instead use the estimated $|B(\xx)|$ value to counteract the rapid growth (or decay) of $\tilde{\Lambda}(\xx)$. A factor of $n^{2\Delta - 5/2}$ is extracted for the same reason. The resulting data is then fitted to the asymptotic form of $n^{-2\Delta + 5/2}|B(\xx)|^n \tilde{\Lambda}(\xx)$ (still given by (\ref{eq:small_x_asymp})), producing much more precise estimates of $(\arg{A(\Delta, \xx)}, \arg{B(\xx)})$. The comparison between the fits is shown in figure (\ref{fig:log_osc_fit_compare}).

This technique works for most values of $\xx$; failing for values close to $\xx = 0$ due to the nature of the power series (\ref{eq:small_x_pwr_series}), and also for values approaching $\xx = \xx_c$ since the oscillation period becomes large enough to interfere with amplitude envelope predictions. 
\begin{figure}
    \centering
    \includegraphics[width=0.7\linewidth]{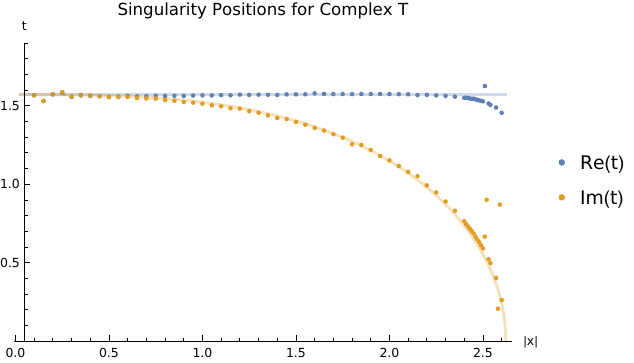}
    \caption{Comparison between the OPE computations and geodesic predictions of singularities occurring for $\xx < \xx_c$, for $\Delta=11/3$. }
    \label{fig:complex_t_sing}
\end{figure}

\begin{figure}
    \centering
    \includegraphics[width=0.45\linewidth]{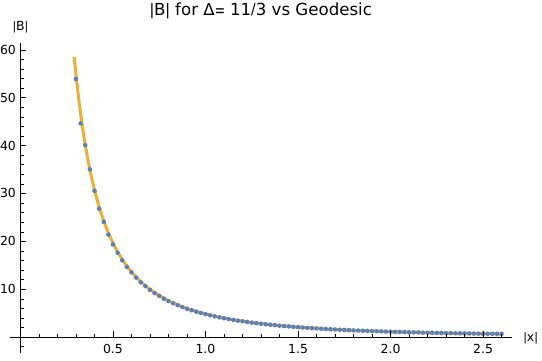}
    \includegraphics[width=0.45\linewidth]{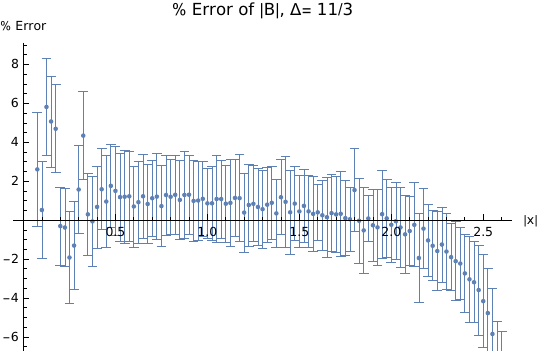}
    \caption{Left: Comparison of the values of $|B(|\vec{x}|)|$ from data fits to the geodesic values. Right: Percentage deviation from geodesic values, including a 95\% confidence interval. As $|x|$ approaches zero, the value of $|B|$ diverges and it is expected that the accuracy of the OPE method decreases significantly.}
    \label{fig:abs_b_plots}
\end{figure}

\begin{figure}
    \centering
    \includegraphics[width=0.45\linewidth]{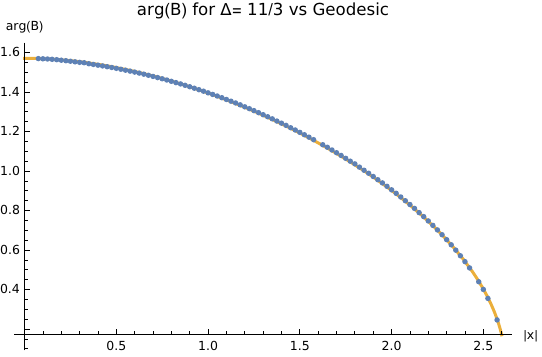}
     \includegraphics[width=0.45\linewidth]{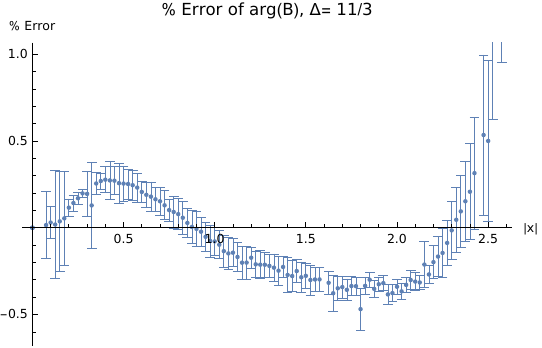}
    \caption{Left: Comparison of the values of $arg(B(|\vec{x}|))$ from data fits to the geodesic values. Right: Percentage deviation from geodesic values, including a 95\% confidence interval. The data used for the fitting procedure is dependent on the estimated $|B|$ value, yet are much more accurate.}
    \label{fig:arg_b_plots}
\end{figure}

The position of the singularity in $t$ as a function of $\xx$ is given by figure (\ref{fig:complex_t_sing}), showing both real and imaginary values. The position smoothly interpolates between the known points of $t=\pi/2, \xx = \xx_c$ and $t=\frac{\pi}{2}+i \frac{\pi}{2}, \xx=0$. To compare the data to analytic predictions, we consider $|B(\xx)|$ and $\arg{B(\xx)}$ separately to highlight the accuracy of the two fitting procedures used. Results of the fitting procedures and the relative errors are shown in figures (\ref{fig:abs_b_plots}) and (\ref{fig:arg_b_plots}). The numerical computation returns values of $|B(|\vec{x}|)|$ that match to the geodesic predictions to within 5\%, and $\arg{B(|\vec{x}|)}$ that match to within 0.5\%.

Consider now the complex geodesics described in \eqref{eq:geo_derv_def} with the momentum parameter $b$ taken to be purely imaginary, $b = \pm i |b|$. The complex turning point at $u=-u_* = -\sqrt{1+|b|^2}$ now exists in the interior of the black hole in the complexified geometry \eqref{eq:exotic_bulk}.
The resulting spatial displacement is real, bounded above by (\ref{eq:Xforbzero}) at $|b|=0$ and approaches $|\vec{X}| = 0$ as $|b| \to \infty$,
\begin{equation}
    |\vec{X}|(\pm i |b|) = \frac{2 \sqrt{\pi} \Gamma(5/4)}{\Gamma(3/4)(1+|b|^2)^{1/4}}\,.
\end{equation}

The expression for time displacement can be split into real and imaginary parts as follows,
\begin{equation}
    T(\pm i |b|) = \frac{\pi}{2} \pm i \frac{\pi}{6} \frac{\sqrt{2\pi}}{\Gamma(3/4)^2}\frac{\left( {}_2F_1\left(-\frac{3}{4}, 1, \frac{1}{2}, -|b|^2 \right)-1 \right)}{|b|(1+|b|^2)^{1/4}}\,.
\end{equation}
The real part of $T(b)$ remains constant, while the imaginary part asymptotes to $\pm i \pi/2$ as $|b| \to \infty$. Recall that we identify lorentzian and euclidean time as,
\begin{equation}
    T(b) \equiv t + i \tau\,,
\end{equation}
such that for generic values of $|b|$, this complex geodesic does not have a clear interpretation in a one-sided or two-sided black brane background.  
As in section \ref{sec:geodesics}, to obtain the full set of values of time displacement we must also consider the analytic continuation around $b=0$, given by,
\begin{equation}
    T_{cont.}(\pm i |b|) = - \frac{\pi}{2} \pm i \frac{\pi}{6} \frac{\sqrt{2\pi}}{\Gamma(3/4)^2}\frac{\left( {}_2F_1\left(-\frac{3}{4}, 1, \frac{1}{2}, -|b|^2 \right)-1 \right)}{|b|(1+|b|^2)^{1/4}} = - T^*(\pm i |b|)\,.
\end{equation}
The complex geodesics can hence be described by,
\begin{equation}
    T(|\vec{X}|) = \frac{\pi}{2} + i \frac{|\vec{X}|^3}{3} \frac{\left( {}_2F_1\left(-\frac{3}{4}, 1, \frac{1}{2}, 1 - \frac{|\vec{X}|_0^4}{|\vec{X}|^4} \right)-1\right) }{\sqrt{|\vec{X}|_0^4 - |\vec{X}|^4}}\,,
\end{equation}
with four-fold symmetry equating the values $\left(T(|\vec{X}|),-T(|\vec{X}|), T^*(|\vec{X}|), -T^*(|\vec{X}|) \right)$.

\section{Small Time Behaviour of Singularity Curve}
\label{app_small_t_coeffs}
\begin{figure}
    \centering
    \includegraphics[width=0.4\linewidth]{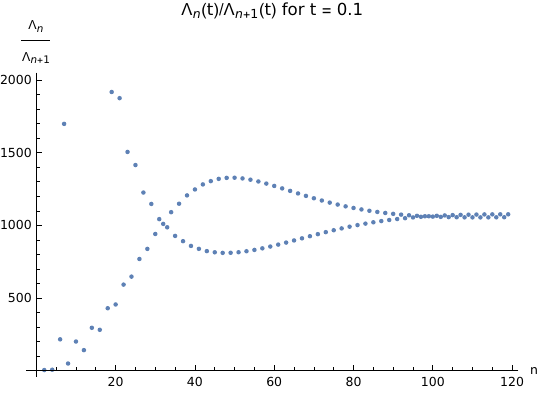}
    \includegraphics[width=0.4\linewidth]{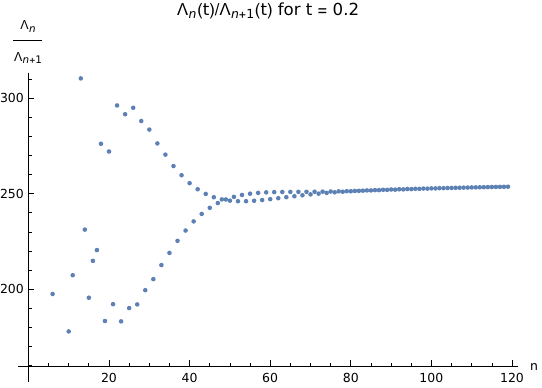}
    \caption{Coefficient convergence for small t}
    \label{fig:t_coeff_convergence}
\end{figure}

In the main text we noted the difficulty in computing the singularity curve $|\vec{x}| = |\vec{x}|(t)$ for small values of t.
The convergence of the coefficients $\Lambda_n$ to the asymptotic form (\ref{eq:t0_sing_ansatz}) is slower for small t. Below $t=0.1$, we do not have enough data points (120) to accurately determine the value of $B(t)$. Plots of the ratio of successive $\Lambda_n$ are shown in figure \eqref{fig:t_coeff_convergence}. The early behaviour (small $n$) of the coefficients for small t is very different from coefficients at large t or for coefficients at $t=0$.

To resolve the above issue, we reorganise the series expansion in \eqref{eq:corr_nonstand_gen} into a power series in $|\vec{x}|$ where $t/|\vec{x}|$ is kept constant. 
\begin{equation}
    \begin{aligned}
         G_T(t,|\vec{x}|) =& \frac{1}{(|\vec{x}|^2-t^2)^{\Delta}} \sum_{i=0}^\infty \sum_{j=0}^i t^{2j} \left( |\vec{x}|^2-t^2\right)^{2i-j} a^i_{j, 2i-j}\\
         =&  \frac{1}{(|\vec{x}|^2-t^2)^{\Delta}} \sum_{i=0}^\infty \sum_{j=0}^i |\vec{x}|^{4i} \left(\frac{t}{|\vec{x}|} \right)^{2j} \left(1-\frac{t^2}{|\vec{x}|^2} \right)^{2i-j} a^i_{j, 2i-j} \\
         =&  \frac{1}{(|\vec{x}|^2-t^2)^{\Delta}} \sum_{n=0}^\infty |\vec{x}|^{4n} \Lambda_n(y) \,, \quad y = \frac{t}{|\vec{x}|}\,,
    \end{aligned}
    \label{eq:GT_series_new_y}
\end{equation}
where we have defined the series coefficients $\Lambda_n(y)$ as follows,
 \begin{equation}
        \Lambda_n(y) = \sum_{j=0}^n y^{2j} \left(1-y^2\right)^{2n-j} a^n_{j, 2n-j} \,.
         \label{eq:Lambda_n_y_coeff_new}
 \end{equation}
Much like in section \ref{sec:bndry_anz_t0}, the asymptotic behaviour of the series coefficients $\Lambda_n(y)$ can identify possible singularities in the $(t, |\vec{x}|)$ space.

The coefficients $\Lambda_n(y)$ approach the same asymptotic form as in \eqref{eq:t_0_asymp_coeff} for large $n$,
 \begin{equation}
   \lim_{n \to \infty}  \Lambda_n (y) = \frac{A(\Delta, y) n^{2 \Delta -5/2}}{B(y)^{4n}}\,,
   \label{eq:Lambda_n_y_coeff_new_asymp}
    \end{equation}
    where $A(\Delta, y)$ is an undetermined function.
We approximate $G_T(t, |\vec{x}|)$ by substituting (\ref{eq:Lambda_n_y_coeff_new_asymp}) into (\ref{eq:corr_lc_seris}) and exchanging the sum with an integral as in (\ref{eq:corr_t0_integral}),
\begin{equation}
\begin{aligned}
      G_T(t, |\vec{x}|) & \approx \frac{A(\Delta, y)}{(|\vec{x}|^2-t^2)^{\Delta}} \int_0^\infty dn \left(\frac{|\vec{x}|^4}{B(y)^4} \right)^n n^{2\Delta - \frac{5}{2}}  \\
      & = \frac{A(\Delta, y) \Gamma\left(2\Delta-\frac{3}{2}\right)}{(|\vec{x}|^2-t^2)^{\Delta}} \left( - \log{\left( \frac{|\vec{x}|^4}{B(y)^4}\right)}\right)^{\frac{3}{2} - 2\Delta}\\
     & \xrightarrow[|\vec{x}| \approx B(y)]{} \frac{A(\Delta, y)}{(1-y^2)^{\Delta}4^{2\Delta-3/2} B(y)^{3/2}}\left(\frac{1}{B(y)-|\vec{x}|}\right)^{2\Delta -\frac{3}{2}}\,.
     \label{eq:GT_singularity_form}
\end{aligned}
\end{equation}
Hence, the correlator exhibits a singularity at,
\begin{equation}
        |\vec{x}| = B(y)\,, \quad t = y B(y)\,.
        \label{eq:corr_sing_line}
\end{equation}

\begin{figure}
    \centering
    \includegraphics[width=0.9\linewidth]{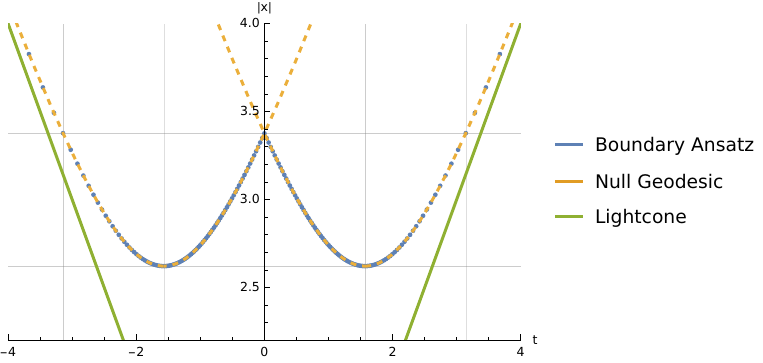}
    \caption{Plot of the position of singularity $|\vec{x}|$ as a function of $t$, as predicted by \eqref{eq:corr_sing_t_val} for $\Delta = 7/3$, alongside the lightcone $|\vec{x}| = t$. Gridlines are shown for $t = \pm \frac{\pi}{2}, \pm \pi$ and $|\vec{x}| = \frac{\sqrt{\pi}}{2}\frac{\Gamma(1/4)}{\Gamma(3/4)}, 3.37$. The upper $|\vec{x}|$ gridline is the singularity position in $|\vec{x}|$ when $t=0$. The gridlines show that the singularity curve is symmetric about $t = \frac{\pi}{2}$. }
    \label{fig:sing_geo_comp}
\end{figure}

The the singularity curve $|\vec{x}| = |\vec{x}|(t)$ is depicted in figure \ref{fig:sing_geo_comp}. To generate the plot, we extract $B(y)$ from $\Lambda_n(y)$ with largest $n$ for a given value of $y$. This allows us to quickly plot upwards of $200$ values of $y$.  Alternatively, one could fit the coefficients $\Lambda_n(y)$ for large $n$ as in figure \ref{fig:sing_geo_comp} for specific values of t. This would result in a more accurate estimate of $B(y)$, however the difference is very small.

The convergence of $\Lambda_n(y)$ to the asymptotic form \eqref{eq:t0_sing_ansatz} is similar to that of the $t=0$ coefficients $a^n_{0, 2n}$ for all values of $y \in (-1, 1)$.  
The series expansion of the correlator \eqref{eq:corr_sing_t0} is centered around $|\vec{x}| = 0$, and requires $y=t/|\vec{x}|$ to be constant. Hence, for each value of $y$ we can investigate how the series \ref{eq:corr_sing_t0} converges along the line given by $t = y |\vec{x}|$. Furthermore, \eqref{eq:corr_sing_t_val} predicts only the first instance away from the origin $t=0, |\vec{x}|=0$ where the series diverges along this line, so the corresponding singularity curve in $(t, |\vec{x}|)$ can only be a single-valued function in polar coordinates.\footnote{For example, this method would not be able to reproduce the singularity curve if, for $t>0$, the slope of the curve exceeds $1$.}
As we vary $y \in (-1, 1)$, the line along which we investigate convergence sweeps the spacelike region of $(t, |\vec{x}|)$. As $y$ approaches the limiting values of $y=\pm 1$ the singularity position $B(y)$ diverges, which indicates that the singularity curve asymptotes to the lightcone $t = \pm |\vec{x}|$ from above. Therefore we also predict that the correlator remains analytic between the lightcone and the singularity curve \eqref{eq:corr_sing_t_val}. However, we cannot comment on the non-analyticities of $G_T(t, |\vec{x}|)$ beyond this region.

\section{Spacelike Geodesics Approaching Null Geodesics}
\label{app_spacelike_geo}

In the previous section, we saw that the positions of the singularities computed at finite $\Delta$ coincide exactly with the null complex geodesics in the asymptotically $AdS_5$ black brane background. 
When the the conformal dimension of the dual field theory operator, $\Delta$, is large, then the correlation function can be approximated by summing over classical saddles of the relevant path integral,
\begin{equation}
    G(t, |x|) \sim \sum e^{-\Delta \mathcal{L}}\,.
    \label{eq:corr_sum_saddles}
\end{equation}
These saddles correspond to the geodesics in the black-hole background that connect the boundary points at which the operators are inserted and $\mathcal{L}$ denotes the regularised proper length of the geodesic. In the limit where spacelike geodesics approach null geodesics, the regularised length diverges and approaches $-\infty$.

In this section we show that complex spacelike geodesics approach null geodesics in the following limit,
\begin{equation}
    |\vec{P}| \to \infty\,, \quad b = \frac{E}{|\vec{P}|}\,, \quad b^2<1\,. 
    \label{eq:space_null_lim}
\end{equation}

The equation for a spacelike geodesic is given by,
\begin{equation}
  \dot{u}^2 = 4 u^2 |\vec{P}|^2 \left(  \frac{1}{|\vec{P}|^2}(1-u^2) -u(1-u^2) + b^2 u \right)\,.
  \label{eq:space_geo_eq}
\end{equation}
We label the solutions to the equation $\dot{u}$, ie. the turning points, by $(u_1, u_2, u_3)$, where $u_1, u_2$ are either both real and positive or complex conjugates, and $u_3$ is the real and negative turning point. 

In the limit given by (\ref{eq:space_null_lim}) the turning points behave as follows,
\begin{equation}
    \begin{aligned}
        u_1 & = \frac{b^2}{(1-b^2)|\vec{P}|^2}+ \mathcal{O}\left( \frac{1}{|\vec{P}|^{4}}\right)\\
        u_2 &=\sqrt{1-b^2} - \frac{b^2}{2(1-b^2)|\vec{P}|^2}+ \mathcal{O}\left( \frac{1}{|\vec{P}|^{4}}\right) \\
        u_3 &= -\sqrt{1-b^2} - \frac{b^2}{2(1-b^2)|\vec{P}|^2}+ \mathcal{O}\left( \frac{1}{|\vec{P}|^{4}}\right)\,.
    \end{aligned}
    \label{eq:turning_point_null_limit_approx}
\end{equation}
The complex spacelike geodesic we will be considering is defined by the turning point $u_3$. The spatial displacement for this complex geodesic is given by,
\begin{equation}
    |\vec{X}|^{(3)}_{s} = \int_0^{u_3} du \frac{u}{\sqrt{u^2 (u-u_1)(u-u_2)(u-u_3)}}\,,
    \label{eq:def_X-integral}
\end{equation}
This can be evaluated in terms of the roots as follows,
\begin{equation}
    \begin{aligned}
        |\vec{X}|^{(3)}_{s} = &  \int_0^{u_3} \frac{du}{\sqrt{(u-u_1)(u-u_2)(u-u_3)}}\\
         =& 2 \sqrt{-u_3}\int_0^1 \frac{dy}{\sqrt{(u_3 - u_1 - u_3 y^2)(u_3-u_2-u_3 y^2)}}\\
         = &\frac{2}{ \sqrt{u_2-u_3}}\int_0^{\sqrt{\frac{-u_3}{u_1-u_3}}} \frac{dt}{\sqrt{1-t^2} \sqrt{1-\frac{u_1-u_3}{u_2-u_3}t^2}}\\
         =& \frac{2}{\sqrt{u_2-u_3}} F\left(\arcsin{\sqrt{\frac{-u_3}{u_1-u_3}}}, \frac{u_1-u_3}{u_2-u_3} \right)\,.
         \label{eq:space_x_u3}
    \end{aligned}
\end{equation}

The time displacement for the complex spacelike geodesic is given by,
\begin{equation}
   T^{(3)}_s = b \int_0^{u_3} du \frac{u}{(1-u^2)\sqrt{u^2 (u-u_1)(u-u_2)(u-u_3)}}
   \label{eq:def_T_integral}
\end{equation}
We can evaluate the time displacement by using partial fractions to express $T^{(3)}_s$ as the sum of the following integrals,
\begin{equation}
    \begin{aligned}
        \frac{b}{2} \int_0^{u_3} \frac{du}{(1 \pm u)\sqrt{(u-u_1)(u-u_2)(u-u_3)}}\,,
        \label{eq:T_partial_sum_dec_}
    \end{aligned}
\end{equation}
and perform the same steps as for $|\vec{X}|^{(3)}_s$ to write them in terms of a new integration variable to obtain,
\begin{equation}
     \frac{b}{ \sqrt{u_2-u_3}(1\pm u_3)}\int_0^{\sqrt{\frac{-u_3}{u_1-u_3}}} \frac{dt}{\left(1- \frac{\mp (u_1-u_3)}{1 \pm u_3}t^2 \right)\sqrt{1-t^2}\sqrt{1-\frac{u_1-u_3}{u_2-u_3}t^2}}\,.
     \label{eq:T_integral_new_coord}
\end{equation}
Combining the above, $T^{(3)}_s$ can be expressed in terms of incomplete elliptic functions of the third kind,
\begin{equation}
\begin{aligned}
      T^{(3)}_s = &\frac{b}{\sqrt{u_2-u_3}} \Bigg( \frac{1}{1+u_3} \Pi \left(\frac{u_1-u_3}{-u_3 - 1}, \arcsin{\sqrt{\frac{-u_3}{u_1-u_3}}}, \frac{u_1-u_3}{u_2-u_3} \right) \\
      & + \frac{1}{1-u_3}\Pi\left(\frac{u_1-u_3}{1-u_3},\arcsin{\sqrt{\frac{-u_3}{u_1-u_3}}}, \frac{u_1-u_3}{u_2-u_3} \right)\Bigg)\,.
\end{aligned}
\label{eq:space_t_u3}
\end{equation}

The integral for the complex geodesic length is given by,
\begin{equation}
    \mathcal{L}^{(3)} =\lim_{\epsilon \to 0^-} \frac{1}{|\vec{P}|} \int_{\epsilon}^{u_3} \frac{du}{u \sqrt{(u-u_1)(u-u_2)(u-u_3)}} + \log{\epsilon}\,,
\end{equation}
where we have regulated the conformal boundary using $\epsilon$. The geodesic length is also defined up to an arbitrary constant, which we will disregard.
The integral evaluates to,
\begin{equation}
    \mathcal{L}^{(3)} =\lim_{\epsilon \to 0^-} \frac{2}{|\vec{P}|(-u_3)\sqrt{u_2-u_3}} \Pi\left(\frac{u_1-u_3}{-u_3}, \arcsin{\sqrt{\frac{\epsilon-u_3}{u_1-u_3}}}, \frac{u_1-u_3}{u_2-u_3}\right) + \log{\epsilon}\,.
\end{equation}
Upon applying an identity to the Elliptic function we can isolate the divergent behaviour as $\epsilon \to 0$ to obtain the regulated complex geodesic length,
\begin{equation}
\begin{aligned}
       \mathcal{L}^{(3)} =& \frac{2}{|\vec{P}|(-u_3)\sqrt{u_2-u_3}} \times \left\{F\left(\arcsin{\sqrt{\frac{-u_3}{u_1-u_3}}}, \frac{u_1-u_3}{u_2-u_3} \right)\right. \\  
       &\left.\qquad\qquad\quad -\Pi\left(\frac{-u_3}{u_2-u_3}, \arcsin{\sqrt{\frac{-u_3}{u_1-u_3}}}, \frac{u_1-u_3}{u_2-u_3}\right) \right\} \\
       &  - \log{\left( \frac{1}{u_3}-\frac{1}{u_2}-\frac{1}{u_1}\right)}\,.
       \label{eq:space_l_u3}
\end{aligned}
\end{equation}

The expressions for the remaining geodesics with turning points $(u_1, u_2)$ can be extracted from the expressions of the complex geodesic by permuting the roots $u_1, u_2, u_3$.
The real geodesic with turning point $u_1$ can be found by exchanging $(u_1, u_3)$ in (\ref{eq:space_x_u3}), (\ref{eq:space_t_u3}) and (\ref{eq:space_l_u3}). The resulting expressions characterising the real geodesic are given by,
\begin{equation}
    |\vec{X}|_s^{(1)}= \frac{2}{\sqrt{u_2-u_1}} F\left(\arcsin{\sqrt{\frac{u_1}{u_1-u_3}}}, \frac{u_3-u_1}{u_2-u_1} \right)\,.
    \label{eq:space_x_u1}
\end{equation}
\begin{equation}
\begin{aligned}
      T_s^{(1)} = &\frac{b}{\sqrt{u_2-u_1}} \Bigg( \frac{1}{1+u_1} \Pi \left(\frac{u_3-u_1}{-u_1 - 1}, \arcsin{\sqrt{\frac{u_1}{u_1-u_3}}}, \frac{u_3-u_1}{u_2-u_1} \right) \\
      & + \frac{1}{1-u_1}\Pi\left(\frac{u_3-u_1}{1-u_1},\arcsin{\sqrt{\frac{u_1}{u_1-u_3}}}, \frac{u_3-u_1}{u_2-u_1} \right)\Bigg)\,.
\end{aligned}
\label{eq:space_t_u1}
\end{equation}

\begin{equation}
\begin{aligned}
       \mathcal{L}^{(1)} =& \frac{2}{|\vec{P}|(-u_1)\sqrt{u_2-u_1}} \left\{F\left(\arcsin{\sqrt{\frac{u_1}{u_1-u_3}}}, \frac{u_3-u_1}{u_2-u_1} \right)\right.\\  
       &\left.\qquad\quad\,\, -\,\Pi\left(\frac{-u_1}{u_2-u_1}, \arcsin{\sqrt{\frac{u_1}{u_1-u_3}}}, \frac{u_3-u_1}{u_2-u_1}\right) \right\} \\
       &  - \log{\left( \frac{1}{u_1}-\frac{1}{u_2}-\frac{1}{u_3}\right)}\,.
       \label{eq:space_l_u1}
\end{aligned}
\end{equation}

Upon applying the limit in \eqref{eq:space_null_lim}, the complex geodesic becomes null with finite boundary displacements $(T_s^{(3)}, |\vec{X}|_s^{(3)})$ and $\mathcal{L}^{(3)} \to - 2 \log{|\vec{P}|}$ as $|\vec{P}| \to \infty$. While the real geodesic length diverges in the exact same fashion, $\mathcal{L}^{(1)} \to - 2 \log{|\vec{P}|}$, the boundary displacements become trivial, i.e $(T_s^{(1)},|\vec{X}|_s^{(1)}) \to 0$. Hence, while we can see that the complex geodesic becomes null in this limit, we cannot directly compare it with the behaviour of the real geodesic.  
 
It is difficult, in general, to choose a path in the space of momentum parameters $(b, |\vec{P}|)$ for each geodesic such that their boundary displacements remain equal, in order to compare their geodesic lengths along a path in the $(T, |\vec{X}|)$ plane.

To better illustrate how the length of the complex geodesic diverges in the limit where it approaches the null geodesic and therefore dominates over the real geodesic at that point, we restrict our analysis to the lightcone limit, as defined and extensively discussed in \cite{Parnachev:2020fna, Esper:2023jeq}.

Starting from the spacelike geodesic in \eqref{eq:space_geo_eq}, we define new momenta $K_\pm$ given by,
\begin{equation}
    K_{\pm} =\frac{1}{2} \left(E \pm |\vec{P}|\right)\,,
    \label{eq:Lightcone_momenta_def}
\end{equation}
and change the coordinate $u$ to a new dimensionless coordinate $x = - 4 K_+ K_- u$. The lightcone limit is defined by keeping fixed $\mathcal{K}$ defined as
\begin{equation}
  \mathcal{K} = - K_+^3 K_- \,,
  \label{eq:lightcone_momentum_conserved}
\end{equation}
while taking $K_+ \to 0$. The lightcone geodesic equation in this limit reduces to,
\begin{equation}
    \frac{\dot{x}^2}{4x^2} = 1 + \frac{x^3}{{\mathcal{K}}} - 4 x\,.
\label{eq:lightcone_geod_spacelike_eq}
\end{equation}
It is natural to similarly define a new displacement parameter that remains finite in the lightcone limit, given by,
\begin{equation}
    \mathcal{X} = - (X^+)^3 X^-\,, \quad X^{\pm} = 2\left(T \pm |\vec{X}| \right)\,.
\label{eq:X_lightcone_def}
\end{equation}
The reduced system in the lightcone limit is now an effective 1D problem.

The turning points are labelled by $x_1, x_2, x_3$ with the same identification as before (see paragraph below (\ref{eq:space_geo_eq})), {\it i.e.}, $x_3 < 0$, $0< x_1< x_2$. 
The real geodesic that returns to the boundary with turning point at $x = x_1$ has no lightlike analogue. The displacement for this real geodesic is given in terms of AppellF1 functions by,
\begin{equation}
\begin{aligned}
    &\mathcal{X}^{(x_1)} = \frac{2^8 x_1^4}{\mathcal{K}}F_1\left(1, \frac{1}{2}, \frac{1}{2}, \frac{3}{2}, \frac{x_1}{x_2}, \frac{x_1}{x_3} \right)^3 \times\\
    &\times\left\{F_1\left(1, \frac{1}{2}, \frac{1}{2}, \frac{3}{2}, \frac{x_1}{x_2}, \frac{x_1}{x_3} \right) - \frac{4 x_1^2}{15{\mathcal{K}}}F_1\left(3, \frac{1}{2}, \frac{1}{2}, \frac{7}{2}, \frac{x_1}{x_2}, \frac{x_1}{x_3} \right)\right\}\,.
    \end{aligned}
\end{equation}
The corresponding geodesic length can be described in terms of the incomplete elliptic functions $F(z, m), \ \Pi(n, z, m)$ of the first and third kind respectively. 
\begin{equation}
    \begin{aligned}
        \mathcal{L}^{(x_1)}({\mathcal{K}}) & = \frac{2}{x_1}\sqrt{\frac{{\mathcal{K}}}{x_2-x_1}}\left(F\left(z, m\right)-\Pi\left(n, z, m\right) \right) - \ln{\left(\frac{1}{x_1} -\frac{1}{x_2}-\frac{1}{x_3}\right)} \\
        & - \ln{\left( 16I_1({\mathcal{K}})\left(I_1({\mathcal{K}})-\frac{1}{2{\mathcal{K}}}I_3({\mathcal{K}})\right) \right)} + \ln{\left(- X^+ X^- \right)} + 2 \ln{(2)}\,,
    \end{aligned}
\end{equation}
where we set,
\begin{equation}
   n = \frac{x_1}{x_1-x_2}, \ z = \sin^{-1}{\sqrt{\frac{x_1}{x_1-x_3}}}, \ m = \frac{x_1-x_3}{x_1-x_2}\,.
\end{equation}
Notice that we have regularised the geodesic length, ensuring consistency with the OPE limit \cite{Parnachev:2020fna}. The geodesic length $\mathcal{L}^{(x_1)}$ does not experience branch cuts or singularities as a function of the displacement $\mathcal{X}^{(x_1)}$, for real values of $\mathcal{X}^{(x_1)}$.

Now let us describe the behaviour of the complex geodesic with turning point $x_3$. The expressions for displacement and geodesic length are given respectively by,
\begin{equation}
\begin{aligned}
    &\mathcal{X}^{(x_3)} = \frac{2^8 x_3^4}{\mathcal{K}}F_1\left(1, \frac{1}{2}, \frac{1}{2}, \frac{3}{2}, \frac{x_3}{x_1}, \frac{x_3}{x_2} \right)^3 \times\\
    &\times \left\{F_1\left(1, \frac{1}{2}, \frac{1}{2}, \frac{3}{2}, \frac{x_3}{x_1}, \frac{x_3}{x_2} \right) - \frac{4 x_3^2}{15{\mathcal{K}}}F_1\left(3, \frac{1}{2}, \frac{1}{2}, \frac{7}{2}, \frac{x_3}{x_1}, \frac{x_3}{x_2}\right)\right\}\,,
    \end{aligned}
\end{equation}
\begin{equation}
    \begin{aligned}
        \mathcal{L}^{(x_3)}(\mathcal{K}) & = \frac{2}{-x_3}\sqrt{\frac{\mathcal{K}}{x_2-x_3}}\left(F\left(z, m\right)-\Pi\left(n, z, m\right) \right) - \ln{\left(\frac{1}{x_3} -\frac{1}{x_2}-\frac{1}{x_1}\right)} \\
        & - \ln{\left( 16I^{(x_3)}_1(\mathcal{K})\left(I_1^{(x_3)}(\mathcal{K})-\frac{1}{2\mathcal{K}}I_3^{(x_3)}(\mathcal{K})\right) \right)} + \ln{\left(- X^+ X^- \right)} +c^{(x_3)}\,,
    \end{aligned}
\end{equation}
with $(n,m,z)$ defined below,
\begin{equation}
   n = \frac{-x_3}{x_2-x_3}, \ z = \sin^{-1}{\sqrt{\frac{-x_3}{x_1-x_3}}}, \ m = \frac{x_1-x_3}{x_2-x_3}\,.
\end{equation}
Note that we have left the regularisation constant $c^{(x_3)}$ undetermined since it does not affect our results. 

\begin{figure}
    \centering
    \includegraphics[width=0.7\linewidth]{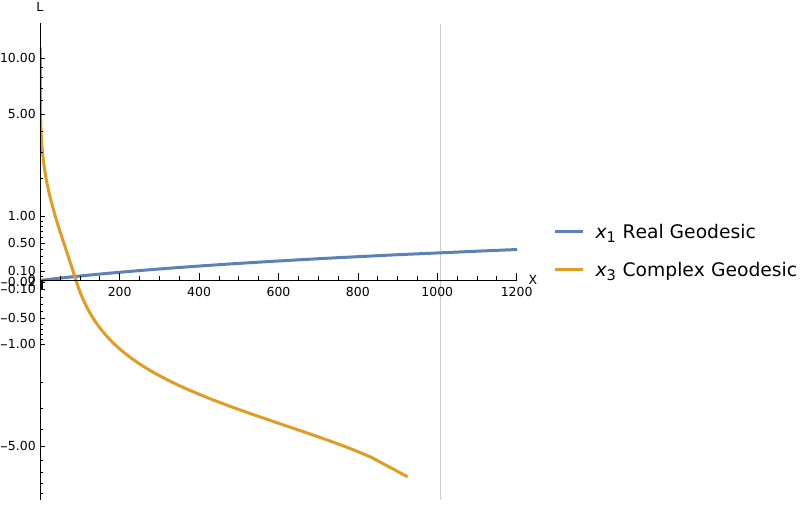}
    \caption{A parametric plot of the real (blue) and complex (orange) spacelike geodesics in the lightcone limit. The real geodesic approaches the origin, from which we can recover the OPE expansion \cite{Parnachev:2020fna}. Since the complex geodesic length approaches positive infinity for small $\mathcal{X}$, it does not contribute to the OPE expansion. The complex geodesic appears more dominant as $\mathcal{X}$ approaches $\mathcal{X} \approx 1008.39$, indicated by the gridline, and ceases to exist for $\mathcal{X}$ greater than this value. The real geodesic exists for all $\mathcal{X}>0$.}
    \label{fig:geo_compare}
\end{figure}

In the limit of $\mathcal{K} \to \infty$, we find the following behaviour,
\begin{equation}
\begin{aligned}
    \mathcal{X}^{(x_3)}\left(\mathcal{K} \to \infty \right) &= \frac{4 \pi^2}{3} \left( \frac{\Gamma(1/4)}{\Gamma(3/4)}\right)^4 \approx 1008.39\,,\\
    \mathcal{L}^{(x_3)}(\mathcal{K} \to \infty) &=  \ln{\left(- X^+ X^- \right)} - \frac{1}{2}\ln{(\mathcal{K})}\,.
    \end{aligned}
\label{eq:x_lc_geo_sing}
\end{equation}
The divergence of the geodesic length signifies that the spacelike geodesic becomes null in this limit, and appears to dominate over the other real geodesic. We compare the real and complex geodesics numerically in figure (\ref{fig:geo_compare}). 
The value of the displacement parameter $\mathcal{X}$ in (\ref{eq:x_lc_geo_sing}) represents the point where the correlator becomes singular in the lightcone limit. This was confirmed numerically, by performing the near boundary analysis of section \ref{sec:boundry_ansatz} directly in the lightcone limit, as well as analytically, by taking the lightcone limit of the displacement associated to the null, complex geodesic (this is a simple exercise given the explicit expressions of sections \ref{sec:geodesics} and \ref{sec:thermal_two_point}).

\section{Geodesic Paths in Complexified Bulk}
\label{app_exotic_bulk_geo}
Curiosity leads us to explore the spacetime in the negative $u$ regime, and to interpret how the geodesics described in section \ref{sec:geodesics} behave in this background. 
The background in the negative $u$ regime is given explicitly the following metric,
\begin{equation}
    ds^2 = \frac{(1-\hat{u}^2)}{\hat{u}}dt^2 - \frac{1}{\hat{u}}d\vec{x}^2 + \frac{1}{4\hat{u}^2 (1-\hat{u}^2)}d\hat{u}^2\,,
    \label{eq:exotic_bulk}
\end{equation}
where we have defined $\hat{u} = -u$ in \eqref{BBAdS5}.
Notice that as $\hat{u}$ approaches the conformal boundary at $\hat{u}=0$, the boundary metric will have the opposite signature with respect to the original metric \eqref{eq:bcknd}, such that the roles of space and time are reversed. The horizon of this complexified bulk is located at $\hat{u} = 1$, and the singularity is at $\hat{u} \to \infty$. 
For a given value of time and spatial displacement $(T, |\vec{X}|)$ on the boundary, the path in the bulk along which the corresponding geodesic travels is defined by the integration contour used in the computation of $(T, |\vec{X}|)$, given by (\ref{eq:null_geo_t_x_def}) for the complex geodesics described in section \ref{sec:geodesics}.  

In the complexified bulk given by (\ref{eq:exotic_bulk}), the complex geodesics begin at the conformal boundary at $\hat{u}=0$ and approach the turning point $\hat{u}_*$, given by $\hat{u}_* = \sqrt{1-b^2}$. 
Depending on the value of $(T, |\vec{X}|)$, the geodesic will trace two different types of paths. For values of $T$ corresponding to the standard expression (\ref{eq:geo_t_1}) the geodesic remains outside the complexified horizon at $\hat{u} = 1$. However, if $T$ corresponds to the analytically continued expression (\ref{eq:geo_t_2}), the geodesic must go past the horizon at $\hat{u}=1$, such that it exists on the second Riemann sheet about $\hat{u}=1$, before reaching $\hat{u}_*$. In both cases the geodesic returns to the conformal boundary in a symmetric fashion. 

\bibliographystyle{JHEP}
\bibliography{refs}
\end{document}

%% file: header.tex
\newcommand{\be}{\begin{equation}}
\newcommand{\ee}{\end{equation}}
\newcommand{\bea}{\begin{eqnarray}}
\newcommand{\eea}{\end{eqnarray}}
\newcommand{\bee}{\begin{enumerate}}
\newcommand{\eee}{\end{enumerate}}
\newcommand{\bei}{\begin{itemize}}
\newcommand{\eei}{\end{itemize}}
\newcommand{\bal}{\begin{equation}\begin{aligned}}
\newcommand{\eal}{\end{aligned}\end{equation}}
\newcommand{\bem}{\left (\begin{matrix}}
\newcommand{\eem}{\end{matrix} \right )}

\newcommand{\alg}[1]{\mathfrak{#1}}

\def\sl2{\alg{sl}(2)}

\numberwithin{equation}{section}
\makeatletter
 \let\old@startsection=\@startsection
 \let\oldl@section=\l@section
 \renewcommand{\@startsection}[6]{\old@startsection{#1}{#2}{#3}{#4}{#5}{#6\mathversion{bold}}}
 \renewcommand{\l@section}[2]{\oldl@section{\mathversion{bold}#1}{#2}}
\makeatother

\renewcommand{\geq}{\geqslant}

\definecolor{grey}{rgb}{0.4,0.4,0.5}
\definecolor{darkgreen}{rgb}{0,0.5,0}
\definecolor{darkred}{rgb}{0.6,0.0,0}
\definecolor{lightbrown}{rgb}{1,0.9,0.8}
\definecolor{brown}{rgb}{0.6,0.3,0.3}
\definecolor{darkblue}{rgb}{0,0,0.5}
\definecolor{darkmagenta}{rgb}{0.5,0,0.5}

\definecolor{TCDblue}{rgb}{0.1,0.3,0.6}

\def\a {\alpha}

\def\vU\varUpsilon
\def\vPi{\varPi}

\def\th{{\tilde h}}

\def\x'{\mathaccent 19 x}
\def\y'{\mathaccent 19 y}
\def\n'{\mathaccent 19 n}

\def\et'{\mathaccent 19 \eta}
\def\th'{\mathaccent 19 \theta}
\def\lam'{\mathaccent 19 \lambda}
\def\varet'{\mathaccent 19 \vartheta}
\def\rh'{\mathaccent 19 \rho}
\def\xb'{\mathaccent 19 {\bar{x}}}

\def\d1{{\dot{1}}}

\def\sun2{SU(N)$\times$SU(N)}

\newcommand{\bpm}{\begin{pmatrix}}
\newcommand{\epm}{\end{pmatrix}}

\newcommand{\bw}{\begin{widetext}}
\newcommand{\ew}{\end{widetext}}

\renewcommand{\bar}[1]{\overline{#1}}
\renewcommand{\tilde}[1]{\widetilde{#1}}
\renewcommand{\hat}[1]{\widehat{#1}}

\def \a{{\alpha}}

\newcommand{\sgn}[1]{\text{sgn}\left(#1\right)}